\documentclass[12pt]{iopart}

\usepackage[separate-uncertainty=true]{siunitx}
\usepackage{graphicx}
\usepackage{xr}
\newcommand{\dd}{\mathrm{d}}

\usepackage{iopams}
\begin{document}

\title[Bistability and oscillations in cooperative microtubule and kinetochore dynamics]{Bistability and oscillations in cooperative microtubule and kinetochore dynamics in the mitotic spindle}

\author{Felix Schwietert and Jan Kierfeld}

\address{Physics Department, TU Dortmund University, 44221 Dortmund, Germany}
\ead{jan.kierfeld@tu-dortmund.de}
\vspace{10pt}
\begin{indented}
\item[]March 2020
\end{indented}

\begin{abstract}
  In the mitotic spindle  microtubules attach to kinetochores via
  catch bonds  during metaphase, and microtubule  depolymerization forces
     give rise to  stochastic chromosome oscillations.
  We investigate the  cooperative stochastic microtubule dynamics
in spindle models consisting of ensembles of parallel microtubules, which  
attach to a kinetochore via elastic linkers.
We include  the dynamic instability of microtubules and 
 forces on microtubules and kinetochores from elastic linkers.
A one-sided model, where an external force acts 
on the kinetochore is solved analytically  employing
a  mean-field approach based on  Fokker-Planck equations.
The solution establishes 
a bistable force-velocity relation of the microtubule 
ensemble in agreement with stochastic simulations. 
We derive constraints on linker stiffness and microtubule number 
for bistability. 
The bistable force-velocity relation of the one-sided spindle model 
gives rise to oscillations in the two-sided model, which 
 can explain stochastic chromosome oscillations in metaphase
(directional instability).
We derive constraints on  linker stiffness 
and microtubule number for metaphase chromosome oscillations.
Including poleward microtubule flux 
into the model we can  provide an explanation for the experimentally observed
suppression of chromosome
oscillations in cells with high poleward flux velocities.
Chromosome oscillations persist in the presence of 
polar ejection forces, however, 
with a  reduced  amplitude and a phase shift between sister kinetochores. 
Moreover, polar ejection forces are necessary to align 
the chromosomes at the spindle equator and stabilize an 
alternating oscillation pattern of the two kinetochores.
Finally, we modify the model such that microtubules can only exert
tensile forces on the kinetochore resulting in a tug-of-war 
between the two microtubule ensembles.
Then, induced microtubule catastrophes after reaching
the kinetochore are necessary to stimulate oscillations.
The model can reproduce 
experimental results for 
kinetochore oscillations in PtK1 cells quantitatively.
\end{abstract}

%
\vspace{2pc}

\submitto{\NJP}
%
%
%

\section{Introduction} 

Proper separation of chromosomes during  mitosis is essential for
the maintenance of life and  achieved by the mitotic spindle, which is
composed of two microtubule (MT) 
asters anchored at the spindle poles. 
The spindle contains three types of MTs classified 
according to their function
\cite{dumont2009}: astral MTs interact with the cell membrane to position the
spindle poles, interpolar MTs interact with MTs from the opposite pole to
maintain spindle length, and,  finally,
 kinetochore MTs link to the chromosomes
via the kinetochores at the centromere and can apply pulling forces via the
linkage.  
The MT-kinetochore bond is a catch bond \cite{akiyoshi2010}, i.e., 
tightening under tension  but 
the molecular nature of the MT-kinetochore link is not exactly known
and a complete mechanistic understanding of the catch bond  is missing
\cite{Santaguida2009,joglekar2009} but probably involves 
Aurora B \cite{Sarangapani2014};
the Ndc80 complexes and Dam1 (in yeast) 
are believed  to play a major role in the 
MT-kinetochore link.
One function of the spindle is to align the chromosomes in the
metaphase plate at the spindle equator.  It has been observed in several
vertebrate cells that chromosomes do not rest during metaphase but exhibit
oscillations along the pole to pole axis known as directional
instability \cite{skibbens1993,waters1996,mitchison1989,ganem2005,magidson2011,wan2012,dumont2012},
whereas in Drosophila embryos and Xenopus eggs a directional instability 
does not occur \cite{maddox2002,desai1998}.  
If present, these oscillations are stochastic and 
on the time scale of minutes, i.e., on a much larger time scale 
than the dynamic instability of single MTs. 
Both  single kinetochores and the
inter-kinetochore distance oscillate; inter-kinetochore or breathing
oscillations occur with twice the frequency of  single kinetochore
oscillations \cite{wan2012}.

A quantitative understanding of the underlying mechanics 
of the MT-kinetochore-chromosome system is still
lacking.  In the past, several theoretical models have been proposed
that reproduce chromosome oscillations
\cite{joglekar2002,civelekoglu2006,shtylla2010,shtylla2011,civelekoglu2013,banigan2015,klemm2018}.
(see table \ref{tab:models} and reviews 
\cite{vladimirou2011,civelekoglu2014})
These models have in common that they simplify to a 
quasi-one-dimensional geometry 
and contain two ensembles of MTs growing from the  two
spindle poles that connect to one chromosome
that is represented by two kinetochores connected by a spring (the cohesin bond).
Kinetochores follow overdamped motion
\cite{civelekoglu2006,shtylla2010,shtylla2011,civelekoglu2013,banigan2015}
or are assumed to reach force balance
instantaneously under the influence of
MT depolymerization and polymerization forces (because the friction
force is small) \cite{joglekar2002,klemm2018}.

Several MT depolymerization and polymerization forces
are included into the models. 
The models neglect explicit spindle pole dynamics but
possibly include poleward MT flux~\cite{civelekoglu2006,banigan2015},
which describes a constant flux of tubulin
from the plus-ends towards the spindle pole
and is probably driven by plus-end directed kinesin-5 motors
pushing overlapping antiparallel interpolar MTs apart
and kinesin-13 proteins that 
depolymerize MTs at the centrosome~\cite{kwok2007}.
The main poleward forces on kinetochores are generated by
depolymerization of MTs which builds up 
and transmits a poleward force via the MT-kinetochore link.
Only in the model of
Civelekoglu-Scholey {\it et al.} \cite{civelekoglu2006}
the main poleward force is generated by MT depolymerization
motors at the spindle poles.
In order to be able to exert poleward pulling forces
 the MT-kinetochore bond needs to remain
intact during depolymerization and ``slide''
with the depolymerizing MT plus end.
The force that can be exerted depends on the
nature of this bond and is high if it is
a  catch bond that tightens under tension \cite{akiyoshi2010}.
All models include
switching between polymerizing and depolymerizing MT states;
in most models this switching is caused by catastrophe
and rescue events 
(dynamic instability \cite{mitchison1984}),
only Shtylla and Keener~\cite{shtylla2010,shtylla2011}
 do not introduce  explicit MT catastrophes
 but catastrophe-like events are triggered by a chemical feedback
 loop if MTs approach the kinetochore. 

The two ensembles of MTs are engaged in a kind of tug-of-war
and exert antagonistic forces via the spring connecting kinetochores:
poleward (P) depolymerization forces
of one ensemble generate an antipoleward (AP) force
on the other kinetochore.
Experiments suggest that kinetochore MTs can only exert
P-directed pulling forces by depolymerization but are not
able to directly exert AP-directed pushing forces on the kinetochore
during polymerization \cite{waters1996,khodjakov1996}.
During directional instability,
the spindle switches between the left and the right ensemble
pulling actively in P-direction by depolymerization 
while the respective other 
ensemble is  passively following in AP-direction by polymerization without
actively pushing. 
Nevertheless, some models have included  AP-directed MT pushing forces
\cite{civelekoglu2006,shtylla2010,shtylla2011,banigan2015,klemm2018}. 
 Antagonistic AP-directed forces on the kinetochores
can also be generated
by polar ejection forces (PEFs); they 
originate from non-kinetochore MTs interacting with the chromosome arms 
via collisions or chromokinesins belonging to the kinesin-4 and 
kinesin-10 families~\cite{mazumdar2005}
and
pushing them thereby towards the spindle equator.
The action of different P- and  AP-directed forces can  move
kinetochores  back and forth and also tense and relax the
inter-kinetochore
cohesin bond.

Models differ in their
assumptions about  the MT-kinetochore link and the mechanism
how MT dynamics is directed by mechanical forces
to give rise to kinetochore and 
inter-kinetochore distance oscillations.
The model
 by Joglekar and Hunt \cite{joglekar2002}  uses the
 thermodynamic Hill sleeve model \cite{hill1985}
 for the MT-kinetochore connection,
which assumes equally spaced rigid linkers
that diffuse on the discrete MT lattice.
Shtylla and Keener~\cite{shtylla2010,shtylla2011}
  combine a continuous
  version of the Hill sleeve model with a negative chemical 
  feedback between
the force at the MT-kinetochore interface and the depolymerization rate.
In Hill sleeve models there is no effect of MT insertion
and, thus, force  onto the MT dynamic instability, i.e.,
on catastrophe and rescue rates. The Hill sleeve
can transmit pulling forces onto the kinetochore up to a
critical force above which
MTs pull out of the sleeve \cite{joglekar2002}, and there is
evidence that the Hill sleeve
exhibits catch-bond-like behavior \cite{ghanti2018}. 
More recent studies show that the kinetochore is not rigid,
as supposed in the Hill sleeve model,
but should be viewed as a flexible framework~\cite{oconnell2012}.
Moreover, Ndc80 fibrils have been suggested as main force
transmitter \cite{mcintosh2008,joglekar2009,powers2009}, 
which motivated Keener and Shtylla to modify their Hill sleeve
 model by replacing the rigid attachment sites
with elastic linkers and allowing for a force feedback onto
MT depolymerization~\cite{keener2014}.
However, sleeve models remain speculative
as electron microscopy has not yet
revealed appropriate structures~\cite{dong2007,mcintosh2013}.
Civelekoglu-Scholey {\it et al.} \cite{civelekoglu2006} proposed
 a model in which MTs and kinetochores are linked by  motor
 proteins (dyneins) walking towards the MT minus end; these links
 have no catch-bond-like behavior.
The links  are assumed to be
   able to transmit
   tension onto MTs that promotes MT rescue.
In\ \cite{klemm2018} no explicit linkers are introduced
but permanent MT-kinetochore links are assumed that can
transmit both pulling and pushing forces onto MTs.
As the exact nature of  MT-kinetochore linking
structures is not known, 
a model  of the MT-kinetochore linkage as a generic elastic
structure seems reasonable,
as  in recent models
where the MTs are linked to the kinetochore via (visco-)elastic
springs \cite{civelekoglu2013,banigan2015}.
The MT-kinetochore bond can be modeled as a catch bond, and 
the elastic linkers also transmit forces back onto
the MT allowing for a force feedback onto
MT dynamics as it has been measured in\ \cite{akiyoshi2010}.

In the model of Shtylla and Keener~\cite{shtylla2010},
MTs that are attached to the same kinetochore
share the force from the cohesin bond equally
and exhibit synchronous dynamics.
The last assumption is contradictory to the experimental observation
that one kinetochore MT ensemble does not 
coherently (de)polymerize but always consists
of a mixture of both states~\cite{vandenbeldt2006,armond2015}.
Klemm {\it et al.}~\cite{klemm2018} take into account this observation
by dividing each MT ensemble into a growing and a shrinking sub-ensemble, 
but also make the strong assumption of equal force sharing
between the MTs within each sub-ensemble.
All other models allow for individual MT dynamics
and for different forces between MTs
depending on the distances of MTs to the kinetochore.

The main mechanism for oscillations differs between models
depending on the main antagonistic  AP-directed force
that switches  a depolymerizing P-directed ensemble
back into AP-directed polymerization.
Switching can
be triggered by the  AP-directed force that the other ensemble
can exert via the cohesin spring during  depolymerization
and by AP-directed PEFs
if 
 MT catastrophes are suppressed or  rescues promoted under tension.
In the model by Joglekar and Hunt \cite{joglekar2002}
AP-directed PEFs are essential for switching.
Civelekoglu-Scholey {\it et
al.} \cite{civelekoglu2006} proposed
 a model in which force is transmitted by motor
 proteins.
 By variation of the model parameters they were able to reproduce a
 wide range of chromosome behavior observed in different cell types.
 In this model, a depolymerizing P-directed ensemble
  switches because tension in the cohesin spring and PEFs
 promote rescue events. 
 A modified  model \cite{civelekoglu2013} 
uses viscoelastic catch  bonds and
accounts for the observation that in PtK1 cells only chromosomes in the center
of the metaphase plate exhibit directional instability~\cite{wan2012}.  They
explain this dichotomy with different distributions of PEFs
at the center and the periphery of the metaphase plate.
In the model by Shtylla and Keener~\cite{shtylla2010,shtylla2011}
MT catastrophe-like events are only triggered by a chemical feedback
such that 
kinetochore  oscillations become coupled to  oscillations of the
 chemical negative feedback system: AP-directed  MT polymerization
 exerts pushing forces onto the kinetochore but 
 triggers switching into a depolymerizing state, and MT depolymerization
 exerts P-directed pulling forces and 
 triggers switching back into a polymerizing state.

Whereas in\ \cite{joglekar2002,civelekoglu2006,civelekoglu2013}
 AP-directed  PEFs are present and in the model by Joglekar and Hunt
 \cite{joglekar2002}
  also essential for realistic
 kinetochore oscillations,
 Banigan {\it et
al.} \cite{banigan2015} presented a minimal model with simple
 elastic linkers and 
neglecting PEFs.  Referring to experiments with budding yeast
kinetochores~\cite{akiyoshi2010}, they modeled MT dynamics with 
force-dependent velocities, catastrophe and rescue rates.
In this model, kinetochore oscillations arise
solely from the collective behavior of attached MTs under force
and the resulting interplay between 
P-directed  depolymerization forces
and AP-directed polymerization forces of the
opposing MT ensembles. 
Force-dependent velocities, catastrophe and rescue rates
are essential to trigger switching of kinetochore motion and
oscillations in this model.
MTs can exert pushing forces such that
it is unclear to what extent the oscillation mechanism
remains functional if pushing forces are absent as suggested
experimentally.
Also the recent model by Klemm {\it et al.}~\cite{klemm2018},
which aims to describe kinetochore dynamics in fission yeast,
does not rely on PEFs. It uses a permanent
MT-kinetochore bond and  oscillations result from the
interplay between MT   depolymerization and polymerization forces
via force-dependence in  MT dynamics;
also in this model MTs can exert pushing forces.
Moreover, the model  makes the strong assumption of  equal force
sharing between all growing or shrinking MTs attached to a kinetochore.
The model also includes kinesin-8 motors that enhance the catastrophe
rate and have a centering effect on the chromosome positions.

\begin{table}[!ht]
\caption{Overview of assumptions of models exhibiting
		stochastic chromosome oscillations.
		In the referred sections we discuss
		how poleward flux, PEFs and the absence of pushing forces
		affect kinetochore dynamics in the model used for this work.}
	\label{tab:models}
	\footnotesize
\begin{tabular}{llllllll}
  \br
              &  linker \ & catch \ & equal \ & force-dep.\ & MT 
  &      & pole- \  \\
 Ref.\ (year) &  model \  & bonds \ & force \ &  MT \       & pushing 
  & PEFs & ward \   \\
              &           &         & sharing & rescue/cat.  \  & forces
  &      & MT flux \\
  \mr
  Joglekar \cite{joglekar2002} (2002)  & Hill sleeve &   & no &  no & no
         & yes  & no  \\
Civelekoglu  \cite{civelekoglu2006} (2006) & motor  & no & no & yes  &  yes
         & yes  & yes \\
Civelekoglu \cite{civelekoglu2013} (2013) &  viscoelastic\ & yes & no & yes & no
         & yes & no \\
Shtylla \cite{shtylla2010,shtylla2011} (2010) & Hill sleeve &  & yes & no & yes
                              & yes & no\\
Banigan  \cite{banigan2015} (2015) &  elastic & yes & no & yes & yes
  & no & yes \\
Klemm  \cite{klemm2018} (2018) &  permanent &  & yes & yes & yes 
   & no  & no \\
  this work & elastic & yes & no & yes   & sec.\ \ref{sec:push}
        &  sec.\ \ref{sec:PEF}  & sec.\ \ref{sec:flux} \\           
  \br
\end{tabular}
\end{table}

  In all Refs.\ \cite{joglekar2002,civelekoglu2006,shtylla2010,shtylla2011,civelekoglu2013,banigan2015,klemm2018}
  a sufficient set of ingredients is given for the respective model
  to exhibit oscillations 
including a specific choice of parameter values.
It is much harder to give   necessary conditions and 
parameter ranges
for oscillations, which means to obtain 
quantitative bounds on model parameters, than to give
a sufficient set of conditions.
This is the aim of the present paper within a model that starts 
from  the minimal model by Banigan {\it et al.} and generalizes
this model in several respects in later sections,
see table \ref{tab:models}.
In this way we discuss the complete inventory of possible
forces acting on the kinetochore and its influence on oscillations.

    It is also difficult to trace the actual mechanism
  leading to oscillations.
An essential part in our quantitative  analysis is a mean-field solution
  of the one-sided  minimal model 
  of Banigan {\it et al.} \cite{banigan2015},
  where a single kinetochore under force
  is connected to one or several MTs that experience
  length-dependent individual loads and feature
  force-dependent dynamics.
  Force-velocity relations for a single kinetochore,
  which is connected to one or several MTs
  have been investigated previously 
based on a sleeve-like MT-kinetochore interface
\cite{shtylla2010,shtylla2011,keener2014,ghanti2018}.
Here, we can derive an  analytical solution of the one-sided
minimal model from a novel mean-field approach.
For this purpose, we start from the Fokker-Planck equations for
the length distribution of the MT-kinetochore linkers.  
The only mean-field approximation is to 
 neglect stochastic velocity fluctuations
 of the  attached kinetochore.
   Our solution clearly shows that the force feedback
   of linkers onto the MT depolymerization dynamics via catch
   (or permanent) bonds is essential for a
   bistable force-velocity relation within the minimal model.
   Moreover, 
the stationary state solution
allows us to quantify the  parameter range for a 
bistability
in the parameter plane of MT-kinetochore linker
stiffness and MT numbers.

By interpreting the force-velocity relation as phase
space diagram for the two-sided model as in\ \cite{banigan2015},
we show that bistability in the one-sided model
is a necessary condition for kinetochore oscillations
in the two-sided model.
Beyond that,
we are able  (1) to quantify an
oscillatory regime, in which kinetochores exhibit directional instability, in
the  parameter plane of linker
stiffness and MT numbers predicting that linkers 
have to be sufficiently stiff; 
(2) to describe kinetochore motion in this
oscillatory regime, calculate frequencies which agree with in vivo
measurements~\cite{wan2012} and to explain frequency doubling of breathing
compared to single kinetochore oscillations; (3) to describe kinetochore motion
in the non-oscillatory regime as fluctuations around a fixed point; 
(4)  to show
that high poleward flux velocities move the system out of the oscillatory
regime and thereby explain why directional instability has been observed in
mitotic vertebrate cells but not in Drosophila embryos and Xenopus eggs;
(5) to show that polar ejection forces reduce 
the amplitude  of oscillations, induce a 
 phase shift between sister kinetochores
and are necessary to align the chromosome at the spindle equator;
(6) to derive as necessary condition for oscillations
that either MTs must be able to apply pushing forces on the kinetochore
or a catastrophe has to be induced with increased 
catastrophe rate when a MT reaches the kinetochore.
 All these results are validated by stochastic simulations;
(7) to provide a set of  model parameters that reproduce 
experimental results for 
kinetochore oscillations in PtK1 cells quantitatively.

In particular,  we quantify 
 lower bounds for linker stiffnesses that allow oscillations,
whose value 
depends on the behavior of MTs growing against the kinetochore. 
If kinetochore MTs can exert pushing forces, we find oscillations for 
linker stiffnesses $>\SI{16}{\pico\newton\per\micro\meter}$;
also if MT  catastrophes are induced  upon reaching the kinetochore, 
we find oscillations in a similar range of linker stiffnesses.
These constraints provide useful 
additional information on MT-kinetochore linkers whose 
molecular nature is not completely unraveled up to now.

\section{Mitotic spindle model}

\begin{figure}[!h]
 \centering
 \includegraphics{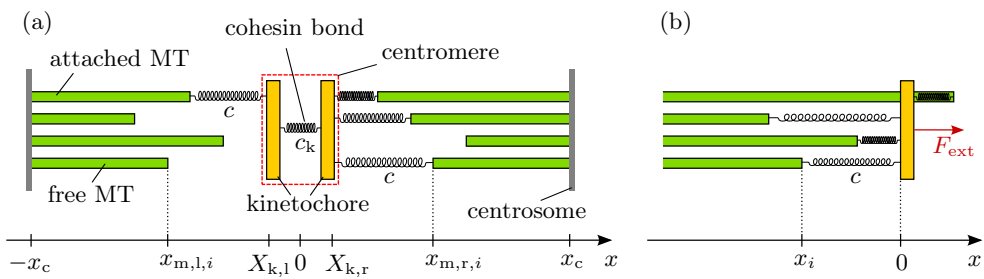}
 \caption{ One-dimensional model of the mitotic spindle. 
     (a)~Two-sided model:
   $M$ MTs arise from each centrosome and can attach to~/~detach from the
   corresponding kinetochore. 
   (b)~One-sided model: Left half of two-sided
   model.  The cohesin bond is replaced by the external force
   $F_\mathrm{ext}$.  MTs are not confined by a centrosome and permanently
   attached to the kinetochore.  MT-kinetochore distances
   $x_i=x_{\mathrm{m,}i}-X_\mathrm{k}$ are the only relevant coordinates.
}
 \label{fig:model}
\end{figure}

We use a one-dimensional model of the mitotic spindle
(figure~\ref{fig:model}(a)),
similar to the model from \cite{banigan2015}.  
The $x$-coordinate specifies positions along the one-dimensional 
model, and we choose $x=0$ to be the spindle equator. 
The spindle model 
contains a single chromosome represented by two kinetochores, which are
linked by a spring with stiffness $c_\mathrm{k}$ and rest length $d_0$.
Two centrosomes margin the spindle at $\pm x_\mathrm{c}$.
From each centrosome a constant number $M$ 
of MTs emerges with their plus ends directed towards the spindle equator. 
Each MT exhibits dynamic instability \cite{mitchison1984} 
and attaches to and detaches from 
the corresponding kinetochore stochastically.  
Attached MTs are connected to the kinetochore by a linker, which we model 
as Hookean polymeric spring with stiffness $c$ and zero rest length. 
This spring exerts a force  $F_\mathrm{mk}  = -c(x_\mathrm{m}-X_\mathrm{k})$ 
on each MT, and each MT exerts a counter force $ -F_\mathrm{mk}$ on the 
kinetochore, where $X_\mathrm{k}$ and $x_\mathrm{m}$ are kinetochore and 
MT position. 

In the following we define all MT parameters for MTs in the left 
half of the spindle model; for MTs in the right half position velocities
$v$ and forces $F$ have opposite signs. 
In the left half, tensile forces  on the MT-kinetochore link 
arise for $X_\mathrm{k} >x_\mathrm{m}$ and  pull the MT in the 
positive $x$-direction,  $F_\mathrm{mk}>0$. 
In \cite{akiyoshi2010}, the velocities of MT growth $v_{\mathrm{m}+}$ and
shrinkage $v_{\mathrm{m}-}$ as well as the rates of catastrophe
$\omega_\mathrm{c}$, rescue $\omega_\mathrm{r}$ and detachment
$\omega_{\mathrm{d}\pm}$ have been measured while MTs were attached 
to reconstituted yeast kinetochores. They can all be described by an 
exponential dependence on  the force $F_\mathrm{mk}$
that acts on the MT plus end:
\begin{eqnarray}
v_{\mathrm{m}\pm} = v^0_\pm \exp\left(\frac{F_\mathrm{mk}}{F_\pm}\right), \qquad
\omega_i = \omega^0_i \exp\left(\frac{F_\mathrm{mk}}{F_i}\right),
\label{eq:expF}
\end{eqnarray}
(for $i=\mathrm{r},\mathrm{c},\mathrm{d}+,\mathrm{d}-$)
with  $F_+,~F_\mathrm{r},~F_\mathrm{d+} >0$ and 
$F_-,~F_\mathrm{c},~F_\mathrm{d-} <0$ for the characteristic
forces, because 
 tension ($F_\mathrm{mk}>0$) enhances growth velocity,
rescue and detachment of a growing MT,
 while it suppresses shrinking velocity, catastrophe 
and detachment of a shrinking MT
(note that we use signed velocities throughout the paper, 
i.e.,  $v_{\mathrm{m}-}<0$ and $v_{\mathrm{m}+}>0$).
Suppression of detachment of shrinking MTs is the catch bond property of the 
MT-kinetochore link. 
The attachment rate is assumed to follow a  Boltzmann distribution,
\begin{eqnarray}
	\omega_\mathrm{a} =   \omega^0_\mathrm{a}
 \exp\left(\frac{c(X_\mathrm{k}-x_\mathrm{m})^2}{2k_\mathrm{B}T}\right),
\label{eq:attach}
\end{eqnarray}
according  to the MT-kinetochore linker spring energy. 

The kinetochore motion is described by an overdamped dynamics, 
\begin{eqnarray}
	v_\mathrm{k} \equiv \dot{X}_\mathrm{k} 
    = \frac{1}{\gamma} \left( F_\mathrm{kk} +  F_\mathrm{km} \right),
\label{eq:vk0}
\end{eqnarray}
with the friction coefficient $\gamma$, and the forces $F_\mathrm{kk}$ and
$F_\mathrm{km}= -\sum_{\rm att.~MTs} F_\mathrm{mk}$ 
 originating from the cohesin bond between kinetochores 
and the MT-kinetochore linkers of all attached MTs,
respectively.

We perform simulations of the model
by integrating the deterministic equations of motion for 
MTs ($\dot{x}_\mathrm{m,i} = v_{\mathrm{m}\pm,i}$ for $i=1,...,M$) 
and kinetochores (eq.\ (\ref{eq:vk0}))  and 
include stochastic switching events between growth and 
shrinking as well as for attachment
and detachment to the kinetochore for each MT. 
For integration we employ an Euler algorithm with a 
fixed time step $\Delta t \leq 10^{-3}\,{\rm s}$ which is small 
enough to ensure $\omega_i \Delta t \ll 1$ for all 
stochastic switching events (see table \ref{tab:parameters}). 
The algorithm is described in the supplementary material in more detail.
We use parameter values from experiments as 
 listed in
table\ \ref{tab:parameters}.

\begin{table}[!ht]
	\caption{Model parameters.}
	\label{tab:parameters}
	\begin{indented}
	\lineup
	\item[]\begin{tabular}{llllr}
		\br
		{\bf Transition parameters} & $\omega_i$ &  $\omega_i^0$ ($\si{s^{-1}}$) & $F_i$ ($\si{pN}$) & \\
		\mr
		catastrophe & $\omega_\mathrm{c}$ 		& $\num{0.0019}$	&	$\num{-2.3}$ & \cite{akiyoshi2010}	\\
		rescue  	& $\omega_\mathrm{r}$ 		& $\num{0.024}$	&	\m$\num{6.4}$ & \cite{akiyoshi2010}	\\
		detachment in growing state & $\omega_{\mathrm{d}+}$	& $\num{0.00011}$	&	\m$\num{3.8}$ & \cite{akiyoshi2010}	\\
		detachment in shrinking state& $\omega_{\mathrm{d}-}$	& $\num{0.035}$	&	$\num{-4.0}$ & \cite{akiyoshi2010}\\
		attachment rate & $\omega_\mathrm{a}$ & $\num{1.0}$ & \multicolumn{2}{r}{estimated}\\
		\br\bs
		{\bf Velocity parameters} & $v_{\mathrm{m}\pm}$ & $v_\pm^0$ ($\si{nm s^{-1}}$) & $F_\pm$ ($\si{pN}$) &  \\
		\mr
		growth		& $v_{\mathrm{m}+}$	& \m\0\0$\num{5.2}$	 &	\m$\num{8.7}$ & \cite{akiyoshi2010}	\\
		shrinking 	& $v_{\mathrm{m}-}$	& $\num{-210.0}$ &	$\num{-3.2}$ & \cite{akiyoshi2010}\\
		\br\bs
		{\bf Other parameters} & Symbol & \multicolumn{1}{l}{Value} & & \\
		\mr
		cohesin bond stiffness & $c_\mathrm{k}$ & $\SI{20}{\pico\newton\per\micro\meter}$ & \multicolumn{2}{r}{estimated}\\
		cohesin bond rest length & $d_0$ & $\SI{1}{\micro\metre}$ & \multicolumn{2}{r}{\cite{waters1996}}\\
		centrosome position & $x_\mathrm{c}$ & $\SI{6.8}{\micro m}$ & \multicolumn{2}{r}{\cite{magidson2011}} \\
		friction coefficient & $\gamma$ & $\SI{1}{\pico\newton\second\per\micro\meter}$ & \multicolumn{2}{r}{estimated} \\
		thermal energy & $k_\mathrm{B}T$ & $\SI{4}{\pico\newton\nano\meter}$ & \multicolumn{2}{r}{estimated} \\
		\br
	\end{tabular}
	\end{indented}
\end{table}

We start with the investigation of the minimal model,
i.e.\ neglecting poleward flux and PEFs
and using the same simple spring model for the MT-kinetochore 
linker as Banigan {\it et al.}
where the MT plus ends are able to ``overtake''
 the kinetochore ($x_\mathrm{m} >X_\mathrm{k}$, again for MTs in the left half
 of the spindle)
and thereby exert pushing forces $F_\mathrm{km}>0$ on the kinetochore 
(which could be interpreted as a compression of the MT-kinetochore linker).
Later, we will generalize the minimal model as described in the introduction,
see table~\ref{tab:models}.
In a first step, we add poleward MT flux,
which describes a constant flux of tubulin
from the plus-ends towards the spindle pole~\cite{kwok2007},
by shifting the MT velocities $v_{\mathrm{m}\pm}$.
PEFs, which push the kinetochore away from the pole \cite{mazumdar2005},
will be included in a second step
as external forces,
which depend on the absolute positions of the kinetochores.
Finally, we will take account of the hypothesis that
MTs are not able to apply pushing forces on the kinetochore~\cite{waters1996,khodjakov1996}
by modifying the model such 
that the growth of a MT is stalled or that the MT 
undergoes a catastrophe when it reaches the kinetochore.

At the centrosome, MTs are confined: 
It is reasonable to assume that 
they undergo a forced rescue and detach from the kinetochore
if they shrink to zero length.
If the mean distance of MTs from the spindle equator is 
sufficiently small, $|\langle x_\mathrm{m} \rangle| \ll |x_\mathrm{c}|$,
we can also consider the MTs as  unconfined 
($|x_\mathrm{c}|\rightarrow \infty$). Then
 both MT and kinetochore dynamics 
solely depend on their relative distances and not on absolute positions,
which simplifies the analysis.

\section{Mean-field theory for bistability in the one-sided model}
\label{sec:MFtheory}

We first  examine the  one-sided
model of Banigan {\it et al.}~\cite{banigan2015},
which only consists of the left half of the two-sided model with an
external force applied to the kinetochore (figure~\ref{fig:model}(b)).  In
simulations of this one-sided spindle model, kinetochore movement exhibits
bistable behavior as a function of the applied force,
i.e., within a certain force range there are two metastable states
for the same external force:
In one state the MTs predominantly grow and the kinetochore velocity is positive
while in the other state the kinetochore has a negative velocity
as a consequence of mainly shrinking MTs.
It depends on the history which of these two states is assumed:
When the system enters the bistable area in consequence of a force change,
the kinetochore velocity will maintain its direction (following 
its current metastable branch) until the force is sufficiently large 
that the system leaves the bistable area again (the current metastable 
branch becomes unstable).
Later we will show  that this hysteresis  of the one-sided model can
explain stochastic chromosome oscillations in metaphase if 
two one-sided models are coupled in the full two-sided model.

In the following, we
present a Fokker-Planck mean-field 
approach that lets us derive bistability analytically
and obtain constraints for the occurrence of bistability.
We obtain a system of Fokker-Planck equations
(FPEs) for the $M$ MT-kinetochore distances
 $x_i\equiv x_{\mathrm{m,}i}-X_\mathrm{k}$ ($i=1,...,M$) 
and 
 decouple the MT dynamics in a mean-field 
approximation, which neglects kinetochore velocity fluctuations. 

We make two assumptions. 
First we assume that all $M$ MTs are
always attached to the kinetochore. 
Because the MT-kinetochore links are
 catch bonds this assumption is equivalent to assuming that these
links are predominantly under tension.
We will check below by comparison with numerical simulations 
 to what extent this assumption can be justified. 
 Secondly, we neglect that  MTs are  confined by a
centrosome.  Then, as mentioned above, the only relevant coordinates are the
relative MT-kinetochore distances $x_i$, which measure 
the extension of the   $i$-th linker.

The MTs are coupled because they attach to the same kinetochore:
each MT experiences a force $F_{\mathrm{mk},i} = -cx_i$ from the elastic
linker to the kinetochore, 
which is under tension (compression) for  $x_i<0$ ($x_i>0$); 
the kinetochore is subject 
to the total counter force $F_\mathrm{km} = c \sum_i x_i$.
Therefore, the kinetochore velocity $v_\mathrm{k}$ is a stochastic 
variable depending on {\it all}
 distances $x_i$, on the one hand, but determines the 
velocities $\dot{x}_i=v_{\text{m}\pm}(x_i)-v_\text{k}$ 
of MTs relative to the kinetochores, on the other hand. 
The equations can be decoupled to a good approximation because 
the one-sided system assumes
a steady state with an approximately 
stationary kinetochore velocity $v_\mathrm{k}$ after a short time 
(rather than, 
for example, a cooperative oscillation as for an MT ensemble
pushing against an elastic barrier \cite{Zelinski2013}).
In our mean-field approximation 
 we then  assume a constant kinetochore velocity 
$v_\mathrm{k} \equiv \langle v_\mathrm{k} \rangle$ and neglect all stochastic 
fluctuations around this mean. This mean value 
is determined by the mean linker extension $\langle x \rangle$  via
\begin{equation}
  v_\mathrm{k} = \frac{1}{\gamma}\left(F_\mathrm{ext}+cM\langle
  x\rangle\right).
\label{eq:vk}
\end{equation}
Fluctuations around the mean value are caused 
by fluctuations of the force $F_\mathrm{km} = c \sum_i x_i$ around its mean 
$\langle F_\mathrm{km} \rangle = M c \langle x \rangle$, which become 
small for large $M$ (following the central limit theorem). 

If $v_\mathrm{k}$ is no longer a stochastic variable, the 
dynamics of the MT-kinetochore extensions $x_i$ decouple.
Then, the probability distribution 
for the  $M$ extensions $x_i$ factorizes into  
$M$ identical factors $p_\pm(x_i,t)$, where $p_\pm(x,t)$
are
the probabilities to find one  particular  MT
 in  the growing ($+$) or  shrinking ($-$) state with a 
MT-kinetochore linker extensions $x$. 
We can derive two FPEs for the dynamical evolution of  $p_\pm(x,t)$,
\begin{eqnarray}
	\partial_t p_+(x,t) &=  
		-\omega_\mathrm{c}(x) p_+(x,t) + \omega_\mathrm{r}(x) p_-(x,t) - \partial_x \big(v_+(x) p_+(x,t)\big), \label{eqn:FPG+}\\
	\partial_t p_-(x,t) &=   
		\phantom{-} \omega_\mathrm{c}(x) p_+(x,t) - \omega_\mathrm{r}(x) p_-(x,t) - \partial_x \big(v_-(x) p_-(x,t)\big), \label{eqn:FPG-}
\end{eqnarray}
where $v_\pm(x)$ denotes the relative velocity of MT and kinetochore, 
\begin{eqnarray}
	v_\pm(x) \equiv v_{\text{m}\pm}(x)-v_\text{k}
   = v_\pm^0\exp\left(-\frac{cx}{F_\pm}\right) - v_\text{k},
\label{eqn:vpm}
\end{eqnarray}
and $\omega_{\mathrm c,r}(x) = \omega_{\mathrm c,r}^0
    \exp\left(-{cx}/{F_{\mathrm c,r}}\right)$.
The velocity
$v_\mathrm{k}$ is no longer stochastic but self-consistently 
determined by (\ref{eq:vk}). 
We note that these FPEs are equivalent to 
 single MT FPEs with position-dependent 
velocities, catastrophe and rescue
rates~\cite{verde1992,dogterom1993,Zelinski2012,Zeitz2014}.

We will now obtain the force-velocity relation of the whole 
MT ensemble by first 
solving the FPEs~\eref{eqn:FPG+} and~\eref{eqn:FPG-} 
in the stationary state $\partial_t p_\pm(x,t)=0$  and then 
calculating the mean linker extension 
 $\langle x\rangle$ for given kinetochore 
velocity $v_\mathrm{k}$ using the stationary 
distribution $p_\pm(x)$.  The
external force that is necessary to move the kinetochore 
with velocity $v_\mathrm{k}$ then follows from (\ref{eq:vk}), 
\begin{equation}
F_\mathrm{ext} = \gamma v_\text{k} - cM\langle x\rangle(v_\mathrm{k}).
\label{eqn:Fextvk}
\end{equation}

The MT-kinetochore distance $x$ is limited to a maximal or a minimal value
$x_\mathrm{max}$ or $x_\mathrm{min}$  for a given kinetochore velocity
 $v_\mathrm{k}> 0$ or $<0$, respectively, see
table\ \ref{tab:boundary}.  
These limiting values are reached if the relative 
MT-kinetochore velocities vanish 
after the linker extension $x$ has adjusted:
First we consider $v_k<0$ and a shrinking MT. 
If we start with a compressed linker ($x > 0$),
  the MT starts to shrink fast,
the compression is reduced
and the linker may get under tension ($x<0$)
because the relative velocity is negative, $\dot{x}=v_-(x)<0$.
The MT-kinetochore distance  $x$ continues to 
decrease   until 
 $\dot{x}= v_-(x_\mathrm{min})=0$ in  (\ref{eqn:vpm}),
where the shrinking velocity of the MTs
is the same as the prescribed kinetochore velocity
($v_{\mathrm{m},-}=v_\mathrm{k}$).
Further shrinking to $x<x_\mathrm{min}$ is not possible but 
distances $x>x_\mathrm{min}$ can always be reached 
if MTs are rescued.
If $v_k<0$ and the MT grows, on the other hand, 
there is no upper bound on $x$,
as the relative velocity $\dot{x}=v_+(x)$ is always positive;
$x$ starts to grow  into the  compressive regime $x>0$
and continues to grow  without  upper bound
(for very large compressive linker extensions, MT growth is suppressed,
but the kinetochore still moves such that  $v_+(x)\approx -v_\mathrm{k}>0$).
Analogously, if $v_k>0$ and MTs grow,  $x$ grows 
until 
$\dot{x}=v_+(x_\mathrm{max})=0$,
and smaller distances can be reached by catastrophe
but there is no lower bound on $x$ for  shrinking MTs.
Linker extensions  $x_\mathrm{max}$  ($x_\mathrm{min}$) are reached 
as stationary states
if  catastrophes (rescues) are suppressed (for example, 
because of large forces), such 
 that  MTs  can grow (shrink) for sufficiently long times. 
If the external force $F_\mathrm{ext}$ is prescribed rather than 
a kinetochore velocity, all MTs reach a stationary state  with the 
same velocity $\tilde{v}_\pm$ given by (\ref{eqn:Fextvk}),
$F_\mathrm{ext} = \gamma \tilde{v}_\pm - cMx_\mathrm{max,min}$.
In this stationary state, both  MT-tips and 
kinetochore move with the same velocity 
\begin{eqnarray}
	\tilde{v}_\pm \equiv 
      \frac{MF_\pm}{\gamma} \,
	W\left( \frac{\gamma v^0_\pm}{MF_\pm} 
   \exp\left( \frac{F_\mathrm{ext}}{M F_\pm} \right) \right),
 \label{eqn:vtilde}
\end{eqnarray}
where $W()$ denotes the Lambert-W function (defined by 
 $x = W(x) e^{W(x)}$).

\begin{table}[!ht]
	\caption{
		Maximal or a minimal value
		$x_\mathrm{max}$ or $x_\mathrm{min}$ of the stationary 
		linker extension distribution $p(x)$ from conditions 
		$v_-(x_\mathrm{min})=0$ and $v_+(x_\mathrm{max})=0$. 
		The distance $x_\mathrm{min}$ ($x_\mathrm{max}$)
                is a function of the prescribed kinetochore velocity $v_\mathrm{k}$ 
                and has to be specified separately depending on the 
	        direction of $v_\mathrm{k}$;
		 $x_\mathrm{min}$ ($x_\mathrm{max}$) is approached if the MTs shrink (grow) for a sufficiently long time.
	}
	\label{tab:boundary}
	\begin{indented}
	\item[]\begin{tabular}{l|ll}
		\br
                         & MT shrinks   & MT grows \\
                \br   
		$v_\mathrm{k}>0$ & $v_-(x) < -v_\mathrm{k} \; {\rm always}$ & $v_+(x) > 0 \; \text{for}\;x<x_\mathrm{max}$ \\
			& $x_\mathrm{min} = -\infty$ & $x_\mathrm{max} = ({F_+}/{c})\ln\left({v^0_+}/{v_\mathrm{k}}\right)$\\
		\mr
		$v_\mathrm{k}<0$ & $v_-(x) < 0 \;\text{for}\;x>x_\mathrm{min}$ & $v_+(x) > v_\mathrm{k} \; {\rm always}$ \\
			& $x_\mathrm{min} = ({F_-}/{c})\ln\left({v^0_-}/{v_\mathrm{k}}\right)$ & $x_\mathrm{max} = \infty$\\
		\mr
		$v_\mathrm{k}=0$ & $v_-(x) < 0 \; {\rm always}$ & $v_+(x) > 0 \; {\rm always}$ \\
			& $x_\mathrm{min} = -\infty$ & $x_\mathrm{max} = \infty$\\
		\br
	\end{tabular}
	\end{indented}
\end{table}

\begin{figure}[!h]
	\centering
	\includegraphics{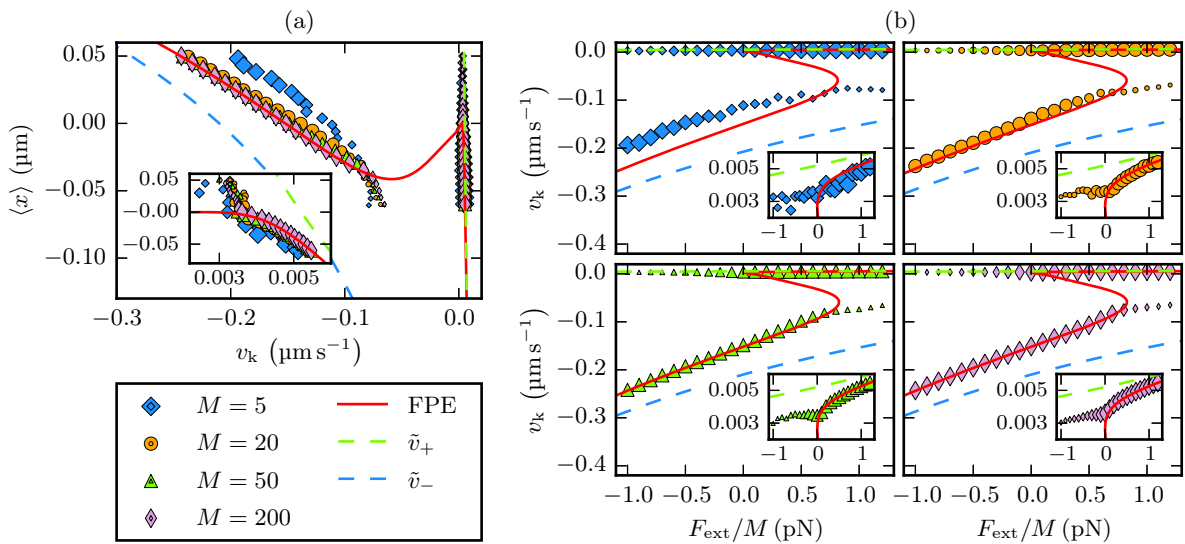}
	\caption{
		 Mean-field results compared to stochastic simulations 
		of the one-sided model.
		(a) The master curve $\langle x\rangle (v_\mathrm{k})$
      from the mean-field approach (red
          line) agrees with simulation results for different MT-numbers
          $M=5,20,50,200$.  
  The  dashed lines mark $x_{\rm min,max}(v_\mathrm{k})$ from table\ \ref{tab:boundary}.
     We run simulations with constant external forces and
          average over 80 simulations for each force.  Initially, the
          MT-kinetochore distance is either $x_\mathrm{min}$ or
          $x_\mathrm{max}$ while all MTs grow or shrink with velocity
          $\tilde{v}_\pm$, respectively.  The system then enters a
          (meta-)stable state, in which we measure
          the mean kinetochore velocity and
          MT-kinetochore distances.  The marker size depicts the time the
          system rests in this state on average, which is a measure for its
          stability (maximum marker size corresponds to 
          $t_\mathrm{rest}\geq\SI{1000}{s}$).  As predicted, the mean-field
          approach turns out to be correct in the limit of many MTs, and in
          this limit the $\langle x\rangle (v_\mathrm{k})$-relation is
          independent of the MT-number $M$. 
       (b) Resulting force-velocity
          relations for different MT-numbers $M=5,20,50,200$.
          The dashed lines  show the large velocity limit $v_\mathrm{k}
\approx \tilde{v}_\pm$ given by \eref{eqn:vtilde}.
         We  used a linker stiffness of
          $c=\SI{20}{\pico\newton\per\micro\meter}$ both in (a) and (b).
}
	\label{fig:xvk}
\end{figure}

In the complete absence of  stochastic switching between growth and 
shrinking by catastrophes and rescues, the MT ensemble reaches
stationary states with peaked distributions
$p_+(x) \propto \delta(x_\mathrm{max}-x)$ and 
$p_-(x) \propto \delta(x-x_\mathrm{min})$.
Stochastic switching shifts and broadens  these peaks, and the FPEs 
\eref{eqn:FPG+} and \eref{eqn:FPG-}  lead to a distribution
$p_\pm(x,t)$
of linker extensions $x$ in the growing and shrinking states
  with statistical weight 
$p_\pm(x,t)>0$  in the whole interval $x_\mathrm{min}\le x\le x_\mathrm{max}$.
At the boundaries 
$x_\mathrm{min}$ and $x_\mathrm{max}$ of this interval, the probability
current density
\begin{eqnarray}
	j(x,t) \equiv v_+(x,t)p_+(x,t) + v_-(x,t)p_-(x,t) 
\label{eqn:jxt}
\end{eqnarray}
has to vanish.  Furthermore, in any  
stationary state ($\partial_t p_\pm(x,t) =0$) 
the current density is homogeneous, as can be seen by summing up
the FPEs \eref{eqn:FPG+} and \eref{eqn:FPG-}:
\begin{eqnarray}
	0 = \partial_x (v_+(x)p_+(x) + v_-(x)p_-(x)) 
    = \partial_x j(x).
 \label{eqn:fluxconst}
\end{eqnarray}
Together with $j=0$ at the boundaries this implies 
 that $j=0$ everywhere  in a steady state.  The resulting
relation $v_+(x)p_+(x) = -v_-(x)p_-(x)$ can be used to reduce the stationary
FPEs to a single  ordinary differential equation with the solution
\cite{Zelinski2012}
\begin{eqnarray}
	p_\pm(x) = \frac{\pm\mathcal{N}}{v_\pm(x)} 
    	\exp\left(- \int\left(\frac{\omega_\text{c}(x)}{v_+(x)}
     +\frac{\omega_\text{r}(x)}{v_-(x)}\right)\dd x \right) 
\label{eqn:p+-}
\end{eqnarray}
for the stationary distribution of linker 
extensions $x$ in the growing and shrinking states.
The normalization constant $\mathcal{N}$ must be chosen so that the overall
probability density $p(x)\equiv p_+(x)+p_-(x)$ satisfies
$\int_{x_\mathrm{min}}^{x_\mathrm{max}} p(x) \dd x = 1$.
Obviously, $p_\pm(x)=0$ for $x>x_\mathrm{max}$ and $x<x_\mathrm{min}$. 
The stationary probability densities $p_\pm(x)$ from \eref{eqn:p+-}
can then be used to calculate the mean distance $\langle x \rangle$ 
as a function of the kinetochore velocity $v_k$, which enters
through \eref{eqn:vpm} for  $v_\pm(x)$.
The integral in the exponent in \eref{eqn:p+-}  as
well as the normalization 
can be evaluated numerically to obtain an explicit 
 $\langle x\rangle (v_\mathrm{k})$-relation, which is  shown in 
figure\ \ref{fig:xvk}(a).

At this point it should be noticed that in the  mean-field theory  the
$\langle x\rangle (v_\mathrm{k})$-relation is {\it independent} of the MT number
$M$.  Therefore, we call it {\it master curve} henceforth. 
 In figure~\ref{fig:xvk}(a) we compare the mean-field theory result to 
stochastic simulations and find that the mean-field approach 
becomes exact in the limit of large $M$, where fluctuations
in the kinetochore velocity  around its mean in (\ref{eq:vk}) 
can be neglected.

The master curve is a central result and will be the basis 
for all  further discussion. Together with the 
force-balance (\ref{eqn:Fextvk}) on the kinetochore,  
the master curve will give the  force-velocity relation 
for the MT-kinetochore system.
A positive slope of the master curve, 
as it can occur for small $v_\mathrm{k}\approx 0$
(see figure\ \ref{fig:xvk}(a)), 
 gives rise to 
an instability of the MT-kinetochore system:
Then, a positive 
kinetochore velocity fluctuation $\delta v_\mathrm{k}>0$ leads to 
a MT-kinetochore  linker compression
 $\delta \langle x\rangle>0$. According to the force-balance 
 (\ref{eqn:Fextvk}),  a compression 
$\delta \langle x\rangle>0$ puts additional forward-force on the kinetochore 
 leading 
to a positive feedback and 
further  increase  $\delta v_\mathrm{k}>0$ of the kinetochore velocity.
This results in an instability, which will prevent the system to assume
mean linker extensions $\langle x \rangle$ in this unstable regime. 
This is confirmed by stochastic simulation results in 
figure\ \ref{fig:xvk}(a), which show that the unstable states are only 
assumed transiently for very short times. 
Therefore, occurrence of 
a positive slope in the master curve in   figure\ \ref{fig:xvk}(a)
is the essential feature that will  
give  rise to bistability in the one-sided model and, finally, 
to oscillations in the full two-sided model.

Now  we want to trace the origin of this 
instability for small  $v_\mathrm{k}\approx 0$. 
If the MTs  are growing (shrinking) for a long time, 
all linker extensions assume
the  stationary values $x\approx x_\mathrm{max}(v_\mathrm{k})$ 
($x\approx x_\mathrm{min}(v_\mathrm{k})$)
from table\ \ref{tab:boundary},
where the  MT-velocity adjusts to the  kinetochore velocity,
$v_\mathrm{k}\approx v_{\text{m}\pm}(x)$. 
If the kinetochore velocity 
 increases in these states by a fluctuation (i.e., $\delta v_\mathrm{k}>0$),
 the MT-kinetochore  linkers are  stretched
(i.e., $\delta x<0$),
which slows  the kinetochore down again
resulting  in  a stable response. 
This is reflected in the negative slopes of 
both   $x_{\rm max}(v_\mathrm{k})$  (for $v_\mathrm{k}>0$) and 
$x_{\rm min}(v_\mathrm{k})$  (for $v_\mathrm{k}<0$).
Because of constant stochastic switching between catastrophes and rescues 
the mean linker extension  exhibits fluctuations about 
$x_{\rm max}$  and $x_{\rm min}$, but we expect also 
the master curve 
 $\langle x\rangle(v_\mathrm{k})$ to 
have a negative slope for a wide range of velocities $v_\mathrm{k}$.
Figure~\ref{fig:xvk}(a)  shows that 
this is actually the case  for kinetochore velocities $v_\mathrm{k}$ 
around the force-free growth or shrinking  velocities $v^0_\pm$ 
of the MTs, i.e., if the imposed kinetochore velocity $v_\mathrm{k}$ 
roughly ``matches'' the force-free growing or shrinking MT velocity.
Then a small  mismatch 
can be accommodated by small  linker extensions $x$, which 
do not dramatically increase fluctuations by 
triggering catastrophe or rescue events.

The situation changes for  small negative or small positive values 
of the kinetochore velocity around $v_\mathrm{k}\approx 0$. 
For $v_\mathrm{k}\lesssim 0$,  MT-kinetochore linkers develop logarithmically 
growing  large negative
extensions  $x_{\rm min}$  (see table\ \ref{tab:boundary}) 
corresponding to  a slow 
  kinetochore trailing  fast shrinking MTs that 
strongly  stretch the linker.
Likewise, for  $v_\mathrm{k}\gtrsim 0$, 
 MT-kinetochore linkers develop logarithmically 
growing  large positive 
extensions  $x_{\rm max}$ corresponding to  a slow 
  kinetochore trailing  fast growing MTs that 
strongly  compress the linker.
Around $v_\mathrm{k}\approx 0$, the system has to switch
from large negative $x$ to large positive $x$ 
because the resulting tensile force $F_{\mathrm{mk}}=-cx$ on the
shrinking  MT will destabilize the shrinking state and give 
rise to MT rescue at least for   $x<-F_r/c$.

Therefore, also the mean value $\langle x\rangle$
switches from negative to positive values 
resulting in  a positive slope of the master curve 
if the stationary distributions $p_-(x)$  and $p_+(x)$  remain sufficiently 
peaked around the linker extensions 
$x_\mathrm{min}$  and $x_\mathrm{max}$, also 
in the presence of fluctuations by catastrophes and rescues.
In the supplementary material, we show that 
the stationary distributions assume a power-law behavior 
$p_+(x) \propto (x_\mathrm{max}-x)^{\alpha_+}$ 
[$p_-(x) \propto (x-x_\mathrm{min})^{\alpha_-}$]
around $x_\mathrm{max}$ [$x_\mathrm{min}$]
for $v_\mathrm{k}>0$ [$v_\mathrm{k}<0$]
with exponents $\alpha_\pm$
that depend on the MT-kinetochore stiffness $c$ as 
$\alpha_\pm +1\propto 1/c$ in the presence of fluctuations.
It follows that distributions are peaked (i.e., 
have a large kurtosis) and bistability emerges
 if the MT-kinetochore linker stiffness $c$ is sufficiently 
large such  that deviations of the MT velocity from the kinetochore 
velocity become suppressed by strong spring forces.
This is one of our main results. 
We also find that 
$\alpha_\pm +1\propto (|v_\mathrm{k}/v_{\pm}^0|)^{-1-|F_\pm/F_\mathrm{c,r}|}$
such that 
the distributions become also peaked around $x_\mathrm{min,max}$ 
in the limit of large velocities $|v_\mathrm{k}|$.
Then the velocity approaches 
$v_\mathrm{k} \approx \tilde{v}_\pm(F_\mathrm{ext})$  for a 
prescribed external force such that 
 $\tilde{v}_\pm$ from (\ref{eqn:vtilde}) represents the large 
velocity and large force limit of the force-velocity relation
of the kinetochore (see figure~\ref{fig:xvk}(b)).

In the unstable regime around 
$v_\mathrm{k}\approx 0$, the linker length distribution $p(x)$ 
is typically broad without pronounced peaks and has a 
minimal kurtosis (as a function of $v_\mathrm{k}$) in the 
presence of catastrophe and rescue fluctuations. In this 
regime the system assumes a state with a heterogeneous stationary 
distribution of growing and shrinking MTs, i.e., 
the total probabilities to grow or shrink become comparable,
$\int p_+(x) \dd x \sim \int p_-(x) \dd x$. If the kinetochore 
velocity is increased, $\delta v_\mathrm{k}>0$,
the system does not react by $\delta x<0$, i.e., by
increasing the average 
tension in the linkers in order to pull MTs forward,
but  by {\it switching} MTs from the shrinking to the growing state
(on average), 
which then even  allows to relax the average linker tension.

Using the force-balance (\ref{eqn:Fextvk}) on the kinetochore,  
the master curve
 is converted to a force-velocity relation for the MT-kinetochore 
system; the results are shown in   figure\ \ref{fig:xvk}(b). 
 The bistability  in the master curve directly translates to a
bistability  in the force-velocity relation of the MT ensemble, and 
 we obtain a 
regime with three branches of 
possible velocities for the same external force.  The upper
and the lower branches agree with our  simulation results and 
previous simulation results in \cite{banigan2015},
and our  mean-field  results
become exact in the limit of large $M$, see figure\ \ref{fig:xvk}(b).
These branches correspond to the two stable parts of the master 
curve with  negative slope, that 
  are found for kinetochore velocities $v_\mathrm{k}$ 
roughly matching the force-free growth or shrinking  velocities $v^0_\pm$ 
of the MTs.
 The mid branch corresponds to 
the part of the master curve with  positive slope, where
the system is unstable. 
Also figure\ \ref{fig:xvk}(b) demonstrates that 
this instability is confirmed by stochastic 
 simulations results.

Finally, we note that a simpler theoretical 
approach, where it is assumed that all linkers 
assume {\it identical} extensions $x_i \approx x$
and all attached MTs are in the same state (growing or shrinking),
is exact for a 
 single MT ($M=1$) by definition but not sufficient
to obtain a bistable force-velocity relation for 
MT ensembles ($M>1$)  (see supplementary material).
The same assumption of identical MT positions 
has already been used to
study an ensemble of MTs that are connected
to the same kinetochore via Hill sleeve like linkers
\cite{shtylla2010,ghanti2018}.
  The model of Klemm {\it et al.}~\cite{klemm2018}
divides each   MT ensemble into a growing and a shrinking sub-ensemble, 
and assumes equal load sharing only 
between  MTs within each sub-ensemble. We can show that, together
with a force-sensitive rescue force, this 
is sufficient to obtain  a bistable force-velocity relation
in  a corresponding
one-sided model.

\section{Bistability gives rise to oscillations in the two-sided model}
\label{sec:osc} 

\begin{figure}[!h]
	\centering
	\includegraphics[width=\textwidth]{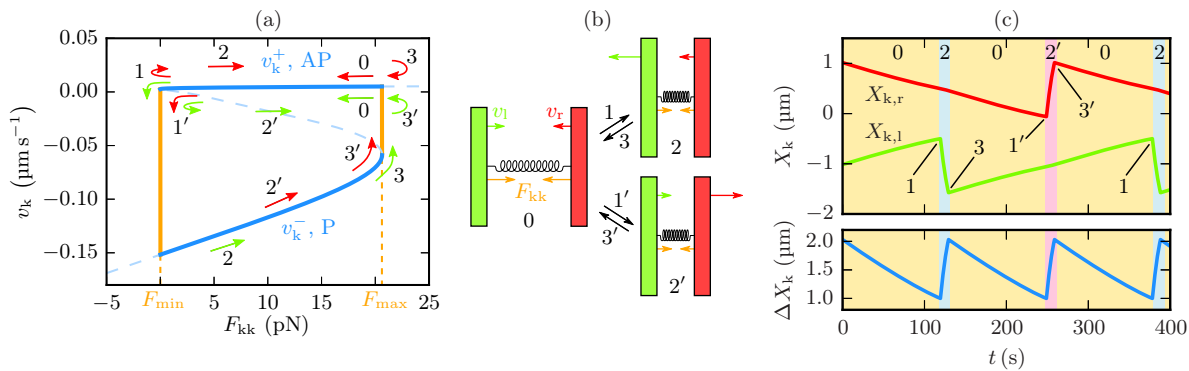}
	\caption{
          Bistability gives rise to oscillations in the two-sided model.
          (a,b) Different states of sister kinetochore motion can be
          deduced from the bistability of the force-velocity relation: either
          both kinetochores are in the upper branch (0) or one is in the upper
          and the other one in the lower branch (2, $2'$).  In the first case,
          both kinetochores move away from their pole (AP) towards each other.
          Thus, the spring force $F_\text{kk}$ decreases until it reaches
          $F_\mathrm{min}$.  Since the upper branch is not stable anymore
          below $F_\mathrm{min}$, either the left (1) or the right ($1'$)
          kinetochore switches to the lower branch and changes direction to
          poleward movement (P).  The system is then in state 2 or $2'$, where
          both kinetochores move into the same direction: the leading
          kinetochore P, the trailing kinetochore AP.  As P- is much
          faster than AP-movement (MT shrinking is much faster than 
          growth), the inter-kinetochore distance and the spring
          force are increasing.  Above $F_\mathrm{max}$ only AP-movement is
          stable, which is why the leading kinetochore changes direction (3,
          $3'$) and the system switches to state 0 again.  
          (c) Solution of the
          equations of motion~\eref{eqn:eom} for
          $c=\SI{20}{\pico\newton\per\micro\meter}$ and $M=25$
          with an imposed periodic order of states ($0-2-0-2'-0-...$).
          The initial condition is $F_\text{kk}=F_\mathrm{max}$ 
          (both kinetochores at the right end of the upper branch).
          For an animated version see video~S1 in the supplementary material.
}
	\label{fig:osc}
\end{figure}

As already worked out by Banigan {\it et al.} \cite{banigan2015}, 
the bistability in the force-velocity relation of the one-sided 
MT ensemble 
can be considered to be the cause for stochastic oscillations in the two-sided
model.
Each ensemble can be either on the lower branch of the 
force-velocity  relation, where it mainly depolymerizes and
exerts a P-directed pulling force ($v_\mathrm{k}<0$)
or on the upper branch where it mainly
polymerizes and exerts an AP-directed pushing force ($v_\mathrm{k}>0$). 
The external force in the one-sided model is a substitute for the
spring force 
$F_\mathrm{kk}=c_\text{k}\left(X_\mathrm{k,r}-X_\mathrm{k,l}-d_0\right)$
 of the cohesin bond in the full model with a 
stiffness $c_\mathrm{k}$ and rest length $d_0$, see 
table\ \ref{tab:parameters}.  Since
the cohesin force is a linear function of the inter-kinetochore distance, the
force-velocity relation can be treated as
distance-velocity (phase space)
diagram for the 
two kinetochores (see figure\ \ref{fig:osc}(a)), where  both 
kinetochores move as points on the force-velocity relation.
 The cohesin
bond always affects the two kinetochores in the same way because 
action equals reaction: if the cohesin spring is
stretched, both kinetochores are pulled away from their pole (AP), if it is
compressed, both kinetochores are pushed polewards (P).  
Thus, the kinetochores
always have the same position on the $F_\mathrm{kk}$-axis in the
$F_\mathrm{kk}$-$v_\mathrm{k}$ diagram in  figure\ \ref{fig:osc}(a),
if  $F_\mathrm{kk}$ on the horizontal axis 
is defined as the  force on the kinetochore in AP-direction
(i.e., $F_\mathrm{kk,l} \equiv F_\mathrm{kk}$ and 
  $F_\mathrm{kk,r} \equiv -F_\mathrm{kk}$  for the left/right kinetochore).  
Likewise, we define $v_\mathrm{k}$ on the vertical axis as the velocity 
in AP-direction (i.e., $v_\mathrm{k,l}\equiv \dot{X}_\mathrm{k,l}$ and 
 $v_\mathrm{k,r}\equiv -\dot{X}_\mathrm{k,r}$ for the left/right kinetochore). 
The upper/lower  stable branch of the 
force-velocity  relation is denoted by  $v^\pm_\mathrm{k}(F_\mathrm{kk})$.
Typically, 
a kinetochore on the upper (lower)
branch has  $v^+_\mathrm{k}>0$ ($v^-_\mathrm{k}<0$) and, thus 
moves in AP-(P-)direction.
Using $F_\mathrm{kk}=c_\text{k}\left(X_\mathrm{k,r}-X_\mathrm{k,l}-d_0\right)$
for the spring force, we find $\dot{F}_\mathrm{kk} = -c_\text{k}
 \left(v_\mathrm{k,r}+v_\mathrm{k,l}\right)$, i.e., 
 kinetochores move with the sum of their AP-velocities along the 
force-velocity curve in the 
$F_\mathrm{kk}$-$v_\mathrm{k}$ diagram.

Oscillations arise from the two kinetochores moving through the
hysteresis loop of the bistable
force-velocity relation as described in figure\ \ref{fig:osc}(a).
Three states are possible (see figure\ \ref{fig:osc}(b)).
In state  $0$, both kinetochores 
move in AP-direction (i.e., in opposite directions) 
relaxing the $F_\mathrm{kk}$-force from the cohesin bond, i.e., 
on the upper branch and to the left in the 
$v_\mathrm{k}$-$F_\mathrm{kk}$-diagram with velocity
 $\dot{F}_\mathrm{kk} = -2c_\text{k} v^+_\mathrm{k}<0$.
After reaching the lower critical force $F_\mathrm{min}$ 
of the hysteresis loop,
  one of the two kinetochores reverses its direction 
 and switches to the lower branch  resulting into states 
$2$ or $2'$ where one kinetochore continues in AP-direction with 
$v^+_\mathrm{k}>0$ while 
the other is moving in P-direction with $v^-_\mathrm{k}<0$
(i.e., both move in the same direction).
In the $v_\mathrm{k}$-$F_\mathrm{kk}$-diagram, 
this results in a motion to the right with velocity
 $\dot{F}_\mathrm{kk} = c_\text{k} (-v^-_\mathrm{k}-v^+_\mathrm{k})>0$
because MTs typically shrink much faster than they grow ($-v^0_-\gg v^0_+$, 
see table\ \ref{tab:parameters}).
Moving on opposite P- and AP-branches increases 
 the kinetochore distance  and builds 
up   $F_\mathrm{kk}$-force  in  the cohesin bond.
After reaching the upper critical force 
$F_\mathrm{max}$ of the hysteresis loop, 
it is always the kinetochore on the lower branch moving in P-direction 
which switches back and state $0$ is reached again. 
This behavior is in agreement with experimental results \cite{wan2012}.
The system oscillates by alternating between state $0$ and one of the states 
$2$ or $2'$  (which is selected randomly with equal probability).

For each of the states 0, 2 and $2'$ depicted in figure\ \ref{fig:osc}(ab) the two
branches $v^\pm_\mathrm{k}=v^\pm_\mathrm{k}[F_\mathrm{kk}]$ provide deterministic
equations of motion for the  kinetochore positions.  Inserting
$F_\mathrm{kk}=c_\text{k}\left(X_\mathrm{k,r}-X_\mathrm{k,l}-d_0\right)$
we obtain both kinetochore  velocities as functions of 
the kinetochore positions and find
\begin{eqnarray}
\eqalign{
    \text{state 0:} \qquad &\dot{X}_\mathrm{k,l} = \phantom{-}v^+_\mathrm{k}\big[c_\text{k}\left(X_\mathrm{k,r}-X_\mathrm{k,l}-d_0\right)\big]>0,\\
					&\dot{X}_\mathrm{k,r} =-v^+_\mathrm{k} \big[c_\text{k}\left(X_\mathrm{k,r}-X_\mathrm{k,l}-d_0\right)\big]<0,\\
    \text{state 2:}  &\dot{X}_\mathrm{k,l} = \phantom{-}v^-_\mathrm{k} \big[c_\text{k}\left(X_\mathrm{k,r}-X_\mathrm{k,l}-d_0\right)\big]<0,\\
                     &\dot{X}_\mathrm{k,r} = -v^+_\mathrm{k} \big[c_\text{k}\left(X_\mathrm{k,r}-X_\mathrm{k,l}-d_0\right)\big]<0,\\
    \text{state $2'$:}  &\dot{X}_\mathrm{k,l} = \phantom{-}v^+_\mathrm{k} \big[c_\text{k}\left(X_\mathrm{k,r}-X_\mathrm{k,l}-d_0\right)\big]>0,\\
						&\dot{X}_\mathrm{k,r} = -v^-_\mathrm{k} \big[c_\text{k}\left(X_\mathrm{k,r}-X_\mathrm{k,l}-d_0\right)\big]>0.
}
	\label{eqn:eom}
\end{eqnarray}
Solving these equations gives idealized deterministic trajectories of the
sister kinetochores, when we also assume that 
  the left and the right kinetochore pass the lower branch 
alternately such that the order of states is a periodic sequence 
 $0-2-0-2'-0-...$ as shown in 
 the  example in figure\ \ref{fig:osc}(c).
 Then single kinetochores oscillate with half the frequency of
 inter-kinetochore (breathing) oscillations, just as observed in
PtK1 cells~\cite{wan2012}.  Moreover, we can obtain numerical values of the
frequencies directly from the trajectories.  For a MT-kinetochore 
linker stiffness
$c=\SI{20}{\pico\newton\per\micro\meter}$ and 20--25 MTs per kinetochore,
which is a realistic number for mammalian cells~\cite{mcewen1997}, we get
periods of 206--$\SI{258}{s}$ and 103--$\SI{129}{s}$ for kinetochore and
breathing oscillations, respectively.  These values coincide with experimental
results of $\SI{239}{s}$ and $\SI{121}{s}$ measured in PtK1
cells~\cite{wan2012}.

\begin{figure}[!h]
	\centering
	\includegraphics[width=\textwidth]{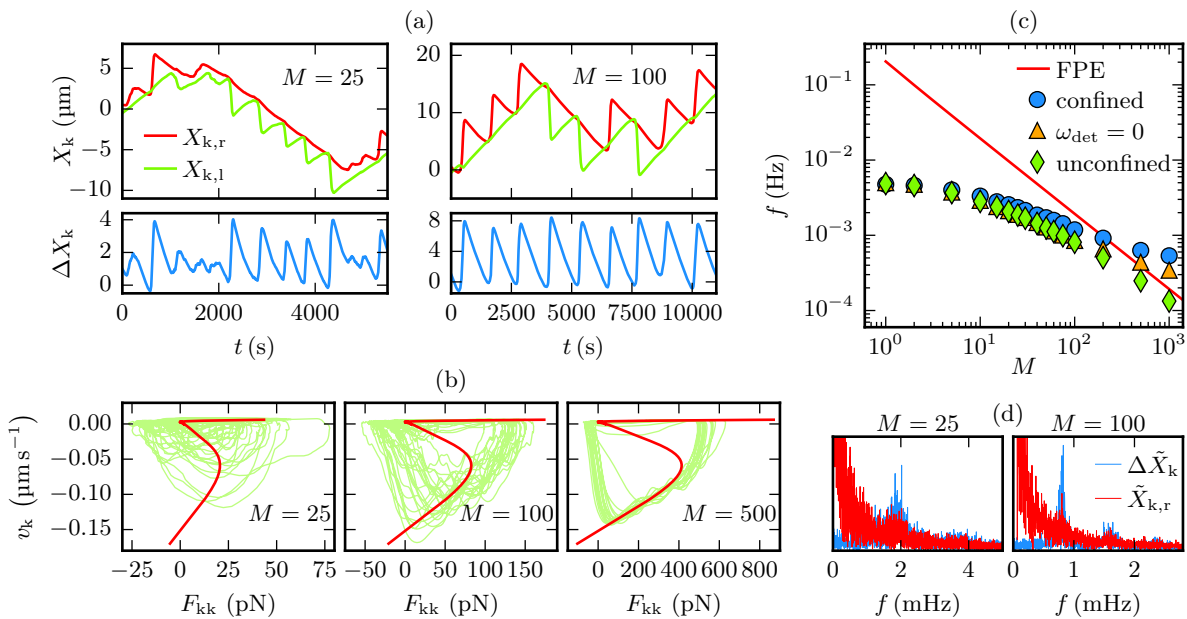}
	\caption{
          Oscillations in stochastic simulations of the unconfined model
          compared to mean-field results.
          (a) Kinetochore trajectories and breathing oscillations in the
          two-sided model without  confinement
          ($x_\mathrm{c}\to\infty$) and 
          detachment ($\omega_\mathrm{d}=0$).
          The kinetochores behave as described in figure~\ref{fig:osc}
          with a random order of states $2/2'$.
          The breathing oscillations are regular enough
          to assign a frequency by Fourier analysis, see (d).
          With less MTs oscillations are more fluctuative.
          (b) Kinetochore velocity against cohesin
          force in simulations of the unconfined two-sided model without
          detachment (green).  For  many MTs 
          the velocity follows very precisely the predicted
          hysteresis from the  mean-field approach (red). 
          For animated versions see videos~S2 ($M=25$)
          and~S3 ($M=500$) in the supplementary material.
          (c) Double-logarithmic plot of frequencies
          of breathing oscillations as a function of MT number $M$: 
          calculated from the mean-field 
          approach  according to figure~\ref{fig:osc} (red)
          and measured in simulations of the
          unconfined (green diamonds) as well as the confined model with
          detachable catch bonds (blue circles)
          and with permanent attachment (orange triangles).
          Confinement becomes relevant for large MT numbers.
          In the presence of detachable catch bonds
          only $\SI{75}{\percent}$ of the MTs
          are attached on average, which  corresponds to a
          simple  shift  of the curve to lower MT numbers.
          (d) Trajectories from (a) in Fourier space.
          While $\tilde{X}_\mathrm{k,r}$ has its maximum at $f=0$
          due to the random order of states in figure~\ref{fig:osc},
          $\Delta\tilde{X}_\mathrm{k}$ has a distinct peak
          that becomes sharper for large $M$
          indicating regular breathing oscillations.
          For all simulations the MT-kinetochore linker stiffness was
          $c=\SI{20}{\pico\newton\per\micro\meter}$.
          }
	\label{fig:unconfined}
\end{figure}

The calculated trajectories are idealized since they
neglect  stochastic fluctuations that occur in simulations of the
two-sided model and have two main effects on the kinetochore dynamics
which already arise in simulations that comply with
the assumptions behind the mean-field theory
(no confinement ($x_\mathrm{c}\to\infty$)
and permanent bonds ($\omega_\mathrm{d}=0$)):
Firstly, the sister kinetochores do not pass the lower branch alternately but
in random order.  Therefore, we observe phases where one kinetochore moves in
AP-direction for several periods, while the other one changes its direction
periodically but moves polewards on average (figure~\ref{fig:unconfined}(a)).
Since this does not influence the trajectory of the inter-kinetochore
distance, breathing oscillations still occur in a more or less regular manner,
which allows us to measure their frequencies by Fourier analysis.
We will show below that additional polar ejection forces 
suppress this random behavior
and force the kinetochores to pass the lower branch alternately.
As a second
effect of the stochastic character of the simulation, kinetochores do not
change the branch instantaneously after crossing the critical forces
$F_\mathrm{max}$ or $F_\mathrm{min}$.  Instead, they tend to maintain their
primary state for a while (figure\ \ref{fig:unconfined}(b)) and follow the
metastable states that we also observe
 in the one-sided model (figure~\ref{fig:xvk}(b)).
Hence, the frequencies we measure in the simulations are smaller than those we
calculate from the Fokker-Planck mean-field 
 approach (figure~\ref{fig:unconfined}(c)).  The
latter effect vanishes in the limit of many MTs (large $M$): 
the switching points
approach the theoretical values $F_\mathrm{max}$ and $F_\mathrm{min}$, and the
simulated breathing frequencies converge to our mean-field predictions.

\begin{figure}[!h]
	\centering
	\includegraphics[width=\textwidth]{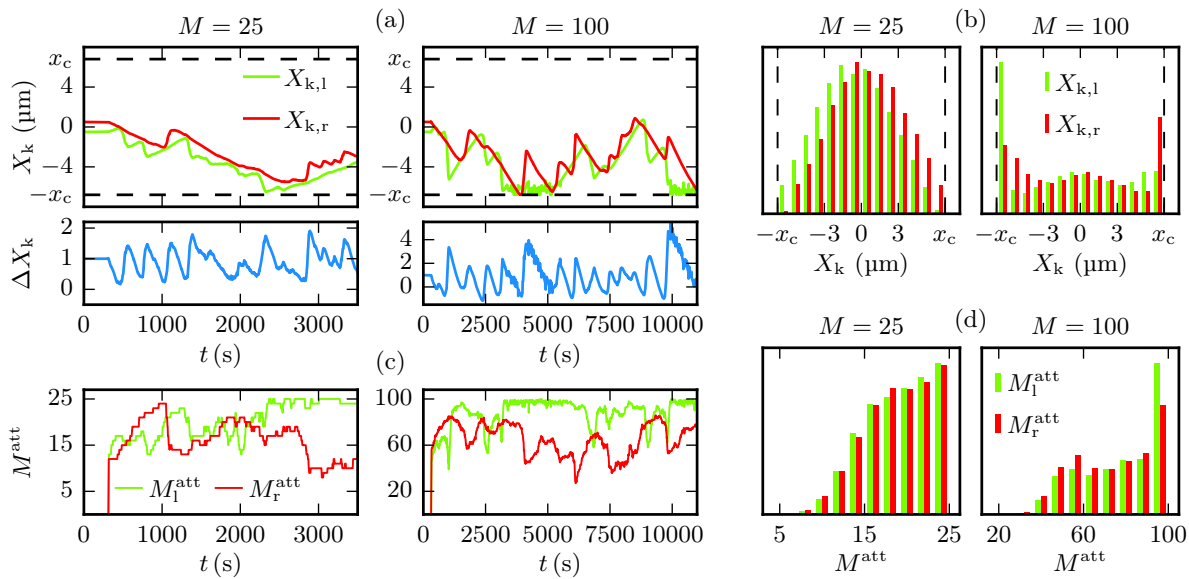}
	\caption{
			Dynamics in the confined model with detachable MTs.
			(a) Kinetochore positions $X_\mathrm{k}$ and 
			inter-kinetochore distance $\Delta X_\mathrm{k}$
			over time
			in simulations with a total number of $M=25$
			and $M=100$ MTs per spindle pole.
			Oscillations as described in figure~\ref{fig:osc}
			are recognizable.
			With 100 MTs one kinetochore can get stuck
			to the centrosome for a while.
			(b) Distribution of kinetochore positions.
			The kinetochores are not aligned to the spindle equator
			and for $M=100$ they are most likely to be found 
			near the centrosomes.
			(c) Number of attached MTs $M^\mathrm{att}$ over time.
			MTs are more likely to be attached when the correspondent
			kinetochore is near the centrosome
			since the free MTs can reattach to the kinetochore
			faster in that case.
			(d) Distribution of $M^\mathrm{att}$. 
			On average $\SI{75}{\percent}$ of the MTs
			are attached independently of the total MT number $M$.
	}
	\label{fig:confined}
\end{figure}

So far we have demonstrated
that the mean field theory correctly describes
kinetochore dynamics in simulations of the unconfined model
where we suppress detachment in order to
prevent unattached MTs from shrinking towards infinity. 
As shown in figure~\ref{fig:confined}(ab),
kinetochore oscillations 
also survive in simulations of the confined model 
independently of whether the MTs are able to
detach from the kinetochore, i.e., to rupture the catch bond.
However,  confinement 
by the centrosome influences the 
kinetochore dynamics in the limit of large $M$:
since more MTs exert a higher force on the kinetochore, 
it is possible that one of
the two sisters gets stuck at  the centrosome for a while
(see figure\ \ref{fig:confined}(ab)).  Hence, the
frequencies measured in the confined two-sided model deviate 
from the frequencies in the unconfined case  above $M\approx
200$.

If we enable detachment in our simulations we
find that the number of attached MTs
correlates with the kinetochore position (see figure~\ref{fig:confined}(c))
since due to the exponential distribution of free MTs
and the distance dependent attachment rate (\ref{eq:attach})
detached MTs are more likely to reattach to the kinetochore
the closer it is to the centrosome.
Moreover, on average, about $\SI{75}{\percent}$ of the MTs are attached
independently of the total MT number (see figure\ \ref{fig:confined}(cd)).
Therefore, the catch bond nature
of the link leads to an effective behavior
similar to a  system without detachment 
but with less MTs,
which explains the difference in frequencies between the confined models with
and without detachment in figure\ \ref{fig:unconfined}(c).
We conclude that detachment does not play a major role
for the occurrence of kinetochore oscillations in cells with many MTs
as despite detachment there are always enough MTs attached
to justify our mean-field approximation.
Hence, (periodic) changes in the number of attached MTs
as they can be seen in figure~\ref{fig:confined}(c)
are rather a passive consequence
than an active source of kinetochore oscillations.
This argumentation may not be tenable,
if just a few MTs are attached to a kinetochore,
so that even detachment of a single MT
effects the total force acting on the kinetochore significantly.
Then, detachment can be the primary cause of directional instability
as worked out by Gay {\it et al.}~\cite{gay2012},
who modeled the mitotic spindle of fission yeast.

Taking into account the results of the last paragraph,
we will mainly investigate 
the unconfined model with permanently attached MTs
in the following sections.
This procedure is reasonable as
we do not lose any qualitative key features of 
kinetochore dynamics on the one hand,
and, on the other hand,
gain a much better comparability 
of our mean field theory with the appropriate stochastic
simulations.

We finally note that 
in all cases we examined
(confined / unconfined system, permanent / detachable bonds)
the kinetochore oscillations become
more fluctuative if less MTs are attached.
This leads to the conclusion
that kinetochore oscillations are a result
of the collective dynamics of an ensemble of MTs
that exhibit a force-dependent dynamic instability individually.
Such a behavior can not be described correctly
based on the simple assumption that all linkers
have the same extension, i.e., that  MTs  share the load equally
and all attached MTs are in the same state (growing or shrinking),
(see supplementary material).
Therefore, the model of Shtylla and Keener~\cite{shtylla2010}
which does assume equal load sharing and synchronous MT dynamics
requires a chemical feedback as an additional mechanism
in order to obtain kinetochore oscillations.
The model of Klemm {\it et al.}~\cite{klemm2018}
divides each MT ensemble into a growing and a shrinking sub-ensemble, 
and assumes equal load sharing only 
between  MTs within each sub-ensemble. Together
with a force-sensitive rescue force, this 
is sufficient to obtain oscillations.

\section{Constraints on linker stiffness and MT number for bistability and oscillations}

\subsection{Constraints for bistability in the one-sided model}

\begin{figure}[!h]
	\centering
	\includegraphics{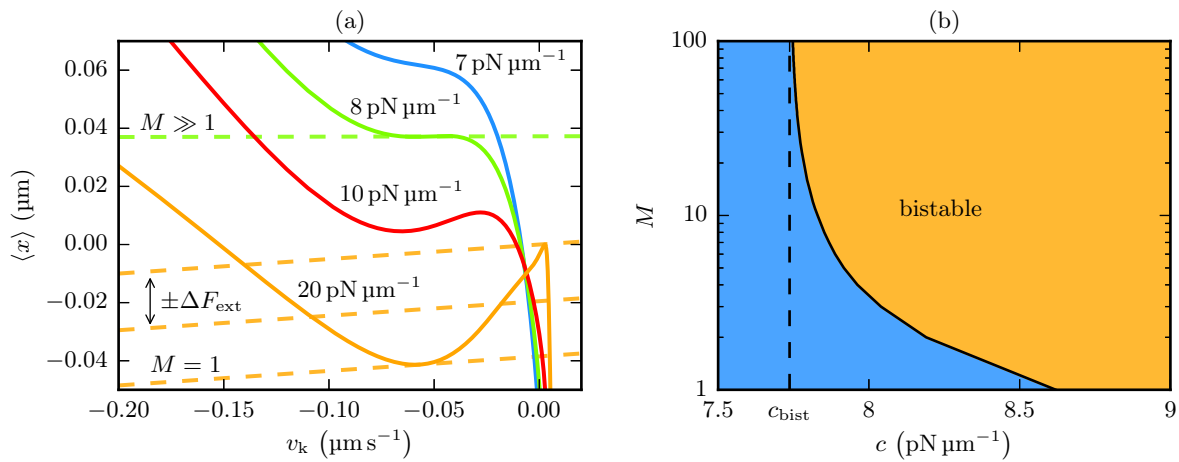}
	\caption{
          Constraints for bistability in the one-sided model.
          (a) Master curves for different linker stiffnesses $c$ and
          linear functions according to~\eref{eqn:xvk_line}.  
          In the limit of large $M$
          the linear functions have zero slope and
          bistability occurs if the master curve has two extrema, which is the
          case for $c>c_\mathrm{bist}$.  For finite $M$ bistable solutions are
          possible if the linear functions have a smaller slope than the
          inflection point of the master curve.  (b) Resulting
          bistable regime in the parameter plane of linker stiffness 
          $c$ and MT number $M$.
  }
\label{fig:phasediagram_bist}
\end{figure}

We already argued above in Sec.\ \ref{sec:MFtheory} 
that bistability (and thus oscillations) can only   emerge
 if the MT-kinetochore linker is sufficiently stiff. 
To analyze the influence of the linker stiffness $c$ and the MT number $M$ 
on  bistability quantitatively, 
the transformation from the master curve to the
force-velocity relation is visualized in figure\ \ref{fig:phasediagram_bist}(a) as
search for the intersections of the  master curve with linear functions
\begin{eqnarray}
\langle x\rangle  = \frac{1}{cM} (\gamma v_\mathrm{k}-F_\mathrm{ext}). 
 \label{eqn:xvk_line}
\end{eqnarray}
In the limit of large $M$ these linear functions have zero slope.  Bistable
force-velocity relations with three intersection
points are only possible if the master curve has  positive slope 
for intermediate $v_\mathrm{k}$ resulting in a maximum and 
minimum.  The extrema of the master curve vanish, however, in a 
saddle-node bifurcation if the linker stiffness drops below 
$c_\mathrm{bist} = \SI{7.737}{\pico\newton\per\micro\meter}$, 
which is, therefore, a lower bound
for the occurrence of bistability.  In the case of finite MT numbers $M$,
bistable force-velocity relations can only be found if the slope in the
inflection point of the master curve exceeds $\gamma/cM$ (the slope of the
linear function \eref{eqn:xvk_line}).  
This allows  us to quantify a
 bistable regime in the parameter plane of linker stiffness $c$ and MT number
$M$  as shown in figure\ \ref{fig:phasediagram_bist}(b).

\subsection{Constraints for oscillations  in the two-sided model}

\begin{figure}[!h]
	\centering
	\includegraphics{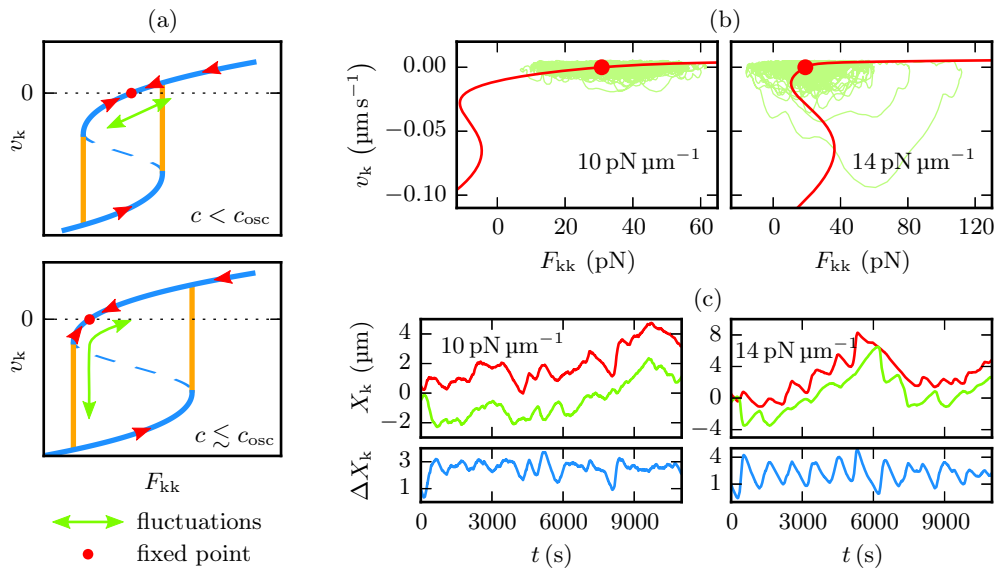}
	\caption{
          Kinetochore dynamics in the non-oscillatory regime.
          (a) Schematic explanation of kinetochore motion in the
          non-oscillatory regime based on  the force-velocity relation.  
          Where the
          upper branch crosses zero velocity, inter-kinetochore distance has a
          fixed point, around which  it fluctuates.  With higher linker
          stiffnesses $c$ the fixed point comes closer to the left turning
          point $F_\mathrm{min}$.  When $c$ is just slightly smaller than
          $c_\mathrm{osc}$, fluctuations can be large enough for the
          kinetochore distance to leave the upper stable branch.  Then, one of
          the two sister kinetochores passes once through 
           the lower branch.  (b,c)
          This behavior can be observed in simulations.  While at
          $c=\SI{10}{\pico\newton\per\micro\meter}$ kinetochores just
          fluctuate around the fixed point, at
          $c=\SI{14}{\pico\newton\per\micro\meter}$ the kinetochores
          occasionally pass through the hysteresis loop. 
           Simulations were performed 
          with an unconfined system and 100 MTs on each side.}
	\label{fig:nonosc}
\end{figure}

We showed in Sec.\ \ref{sec:osc} that 
bistability of the one-sided model is a necessary condition 
for oscillations in the two-sided model. 
Now we show that bistability in the one-sided model is, however, 
{\it  not  sufficient}  for  oscillations in the full model.  
If the
force-velocity relation is interpreted as phase space diagram 
for the two kinetochores, kinetochores only switch branches 
in the $v_\mathrm{k}$-$F_\mathrm{kk}$-diagram 
if  their velocity changes its sign at the turning points
$F_\mathrm{min}$ and $F_\mathrm{max}$.  If  this is not the case and 
 one of the two branches crosses $v_\mathrm{k}=0$
(e.g.\ the right branch for  $c=\SI{10}{\pico\newton\per\micro\meter}$ in
figure~\ref{fig:phasediagram_bist}(a),
which transforms to the upper branch of the force-velocity relation),
the intersection point 
is a stable fixed point in the phase space diagram (see figure\ 
\ref{fig:nonosc}(a)).  
At this fixed point kinetochore motion will relax to zero
velocity and just exhibit
 fluctuations around an equilibrium distance instead of
oscillations.

\begin{figure}[!h]
	\centering
	\includegraphics{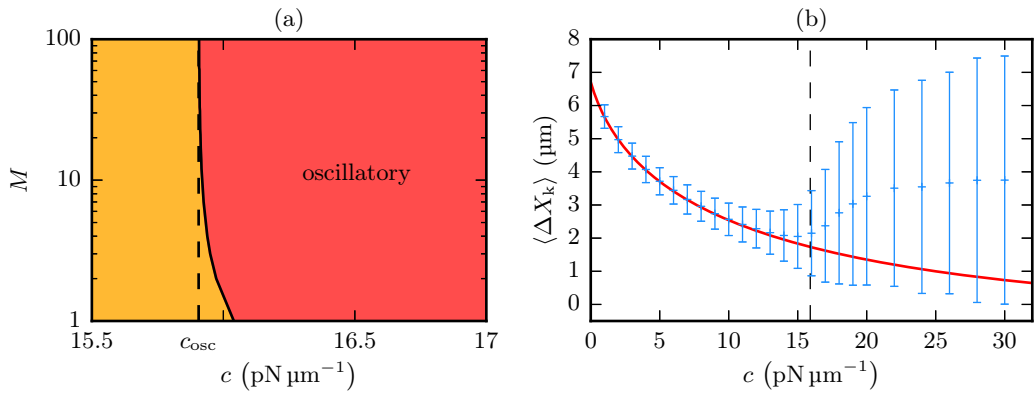}
	\caption{
          Constraints for oscillations in the two-sided model.
          (a) Oscillatory regime in the parameter plane of
          linker stiffness  $c$ and MT number $M$.
          (b) Mean inter-kinetochore distance according to
          \eref{eqn:dxkmean} (red) and measured in simulations (blue)
          with $M=100$.  Below
          $c_\mathrm{osc}=\SI{15.91}{\pico\newton\per\micro\meter}$ 
          (dashed line)
          both results match, whereas in the oscillatory regime mean
          inter-kinetochore distance diverges from the fixed point, and its
          standard deviation increases notably.}
	\label{fig:phasediagram_osc}
\end{figure}

As a sufficient condition for oscillations we have to require -- besides
bistability -- a strictly positive velocity in the upper and a strictly
negative velocity in the lower branch in the 
$v_\mathrm{k}$-$F_\mathrm{kk}$-diagram.  Based on this 
 condition we quantify an
oscillatory regime in the parameter plane of linker stiffness $c$ and MT
number $M$ in figure\ \ref{fig:phasediagram_osc}(a). 
 In the limit of many MTs
the sufficient 
condition for oscillations can be formulated  in terms of the master curve:
 the maximum of the master curve
 has to be  located at a
positive and the minimum at a negative velocity.  This is the case for
$c>c_\mathrm{osc}=\SI{15.91}{\pico\newton\per\micro\meter}$, which is,
therefore, a lower bound for the occurrence of oscillations.
This constraint on the linker stiffness 
for metaphase chromosome oscillations 
provides 
additional information on MT-kinetochore linkers whose 
molecular nature is not known up to now. 

Because of stochastic fluctuations, the  transition between
oscillatory and non-oscillatory  regime is not sharp 
 in our simulations.  In the non-oscillatory regime
kinetochores fluctuate around a fixed point of inter-kinetochore distance,
where  the upper branch crosses $v_\mathrm{k}=0$.  However, these fluctuations
can be large enough for the inter-kinetochore distance to shrink and 
leave the upper
branch on the left side, especially for stiffnesses $c$  slightly below
$c_\mathrm{osc}$.  If that happens, one kinetochore passes once through 
the lower branch
of the force-velocity relation  just as in an oscillation.
  The difference to genuine  oscillations is that these 
are randomly occurring single events (resulting in a Poisson process). 
 Randomly occurring oscillations are visualized in
figure\ \ref{fig:nonosc} for 
$c<c_\mathrm{osc}$ and $c\lesssim c_\mathrm{osc}$.  
Moreover, the force-velocity relations as well as the
kinetochore trajectories measured in 
corresponding simulations are shown.

In the non-oscillatory regime, the fixed point
should determine the mean inter-kinetochore distance 
$\langle\Delta X_\mathrm{k}\rangle = \langle X_\text{k,r} - X_\text{k,l}
\rangle$.  Solving the
FPEs for $v_\text{k}=0$, we compute the (external) force $F_0$ that has to be
applied to  one  kinetochore to stall its motion:
\begin{eqnarray}
	F_0 = \gamma v_\text{k} - cM\langle x\rangle 
       = - cM\langle x\rangle(v_\text{k}=0).
\end{eqnarray}
In the two-sided model this force is applied to the kinetochores by the
cohesin bond at the fixed point.  With $F_\text{kk} = c_\text{k}(\Delta
X_\mathrm{k} - d_0)$ we compute the corresponding 
mean inter-kinetochore distance:
\begin{eqnarray}
 \langle \Delta X_\mathrm{k}\rangle = \frac{F_0}{c_\text{k}} + d_0
     = - \frac{cM}{c_\text{k}}\langle x\rangle(v_\text{k}=0) 
    + d_0. 
\label{eqn:dxkmean}
\end{eqnarray}
Figure~\ref{fig:phasediagram_osc}(b) shows that simulations agree with this result in the
non-oscillatory regime.  At $c_\mathrm{osc}$ the transition to the oscillatory
regime can be recognized, where  the mean inter-kinetochore distance deviates
from the fixed point \eref{eqn:dxkmean}.  Moreover, the variance of 
$\Delta X_\mathrm{k}$ increases significantly at $c_\mathrm{osc}$ 
due to the transition
to the oscillatory regime.

\begin{figure}[!h]
	\centering
	\includegraphics{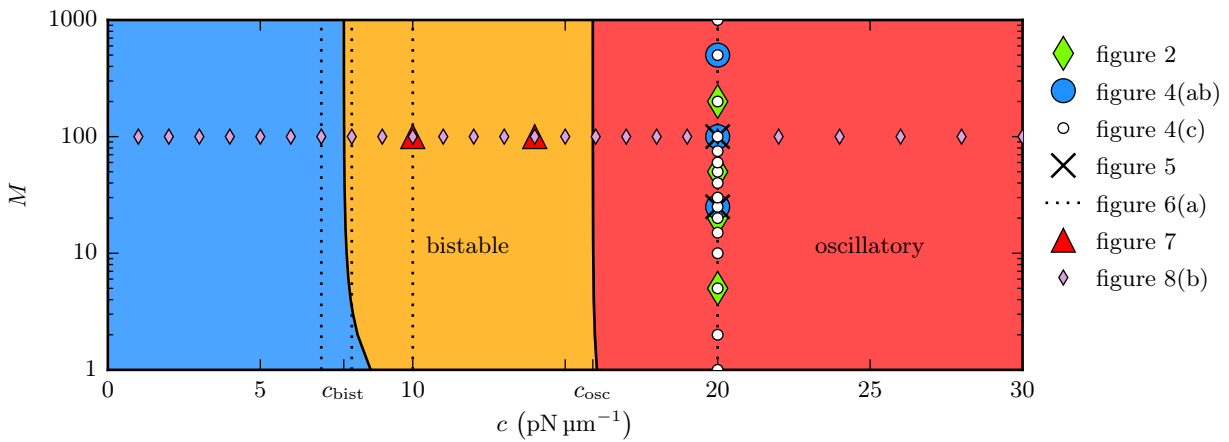}
	\caption{
		Locations in $c$-$M$-parameter plane
		of the master curves from figure~\ref{fig:phasediagram_bist}(a)
		and the simulations from  figures~\ref{fig:xvk}, \ref{fig:unconfined},
		\ref{fig:confined}, \ref{fig:nonosc} and \ref{fig:phasediagram_osc}.
	}
	\label{fig:phasediagram_summary}
\end{figure}

In order to provide an overview
and to make orientation easier for the reader,
we summarize in figure~\ref{fig:phasediagram_summary}
where the stochastic simulations from the last three sections
and the master curves in figure~\ref{fig:phasediagram_bist}(a)
are located in the parameter plane of 
linker stiffness $c$ and MT number $M$,
and which regime they are part of.

\section{Poleward microtubule flux suppresses oscillations}
\label{sec:flux}

An effect we have not included so far is poleward MT flux, which
was observed in several metazoan cells (table\ \ref{tab:flux}). It describes the
constant flux of tubulin from the plus-ends towards the spindle pole and is
probably driven by plus-end directed kinesin-5 motors pushing overlapping
antiparallel MTs apart as well as kinesin-13 proteins that are located at the
centrosome and depolymerize the MTs at their minus-ends~\cite{kwok2007}.
During metaphase, spindle and MT length can be maintained by simultaneous
polymerization at the plus-ends~\cite{rogers2005}, which results in a behavior
similar to  treadmilling of MTs~\cite{margolis1978}.

\begin{table}[!ht]
	\centering
	\caption{
	Metaphase poleward flux velocities $v_\mathrm{f}$ 
    and occurrence of directional instability.
	For a more detailed review of poleward flux measurements
	see \cite{rogers2005}}
	\label{tab:flux}
	\begin{indented}
	\lineup
	\item[]\begin{tabular}{lll}
		\br
		Cell type & $v_\mathrm{f} (\si{\nano\meter\per\second}$)& Directional instability \\
		\mr
		LLC-PK1 (porcine)	& \num{8.3} \cite{mitchison1989} 	& yes \cite{mitchison1989} \\
		PtK1 (rat kangaroo)	& \num{7.7} \cite{zhai1995}			& yes \cite{wan2012} \\
		PtK2 (rat kangaroo)	& \num{10} \cite{mitchison1989}		& yes \cite{dumont2012} \\
		Newt lung 			& \num{9.0} \cite{mitchison1992}	& yes \cite{skibbens1993} \\
		U2OS (human)		& \num{8.8} \cite{ganem2005}		& yes \cite{ganem2005} \\
		\mr
		Drosophila embryo	& \num{32} \cite{brust-mascher2002}	& no \cite{maddox2002} \\
		Xenopus egg			& \num{37} \cite{miyamoto2004}		& no \cite{desai1998} \\
		\br
	\end{tabular}
	\end{indented}
\end{table}

\begin{figure}[!h]
	\centering
	\includegraphics{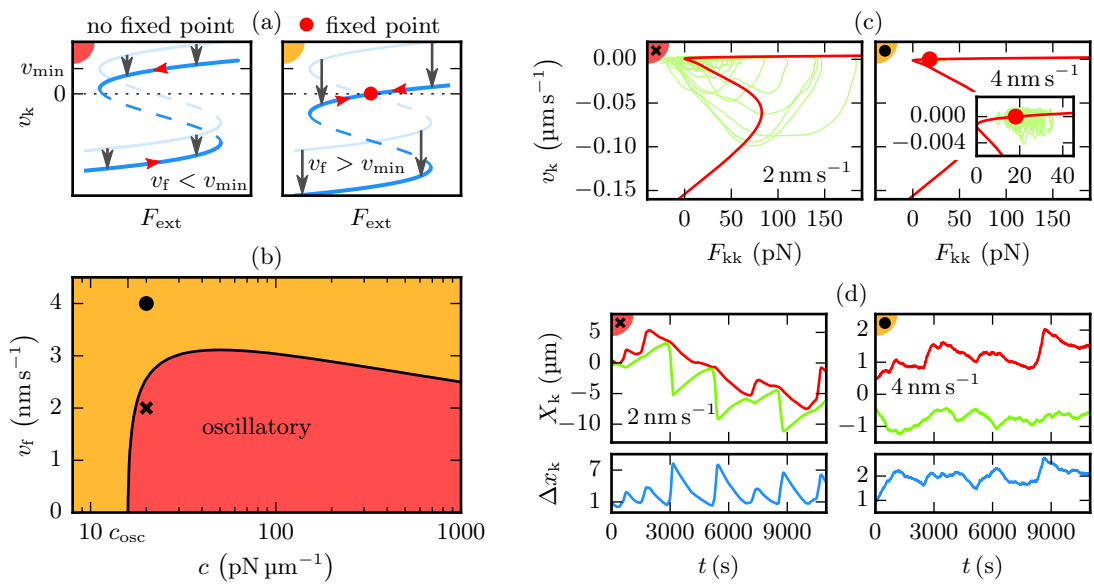}
	\caption{
		Poleward flux suppresses oscillations.
		(a) Due to (\ref{eqn:vpmvf}), the force-velocity relation
		is shifted by the amount of the flux velocity $v_\mathrm{f}$
		towards smaller kinetochore velocities.
		If the flux is slower than the kinetochore velocity $v_\mathrm{min}$
		in the left turning point $F_\mathrm{min}$,
		the kinetochores still oscillate.
		For larger flux velocities,
		a fixed point arises on the upper branch
		and the kinetochores behave as described in figure~\ref{fig:nonosc}.
		(b) Oscillatory regime in the parameter plane
		of $c$ and $v_\mathrm{f}$ in the limit of many MTs.  Fast poleward
		flux suppresses kinetochore oscillations for arbitrary 
		linker stiffnesses $c$. 
		(b,c) Phase space
		diagrams and MT trajectories from simulations of the unconfined
		two-sided model with $c=\SI{20}{\pico\newton\per\micro\meter}$ and
		$M=100$.  While at $v_\mathrm{f}=\SI{2}{\nano\meter\per\second}$ the
		system is still in the oscillatory regime, where hysteresis is
		recognizable in phase space, at
		$v_\mathrm{f}=\SI{4}{\nano\meter\per\second}$ kinetochores show
		fluctuative motion as described in figure~\ref{fig:nonosc}.}
	\label{fig:flux}
\end{figure}

Poleward flux can be easily included in our model by subtracting a constant
flux velocity $v_\mathrm{f}$ from the MT velocity. Then, the relative
MT-kinetochore velocities \eref{eqn:vpm} become
\begin{eqnarray}
	v_\pm(x) = 
    v_\pm^0\exp\left(-\frac{cx}{F_\pm}\right) - v_\mathrm{f} - v_\mathrm{k}.
\label{eqn:vpmvf}
\end{eqnarray}
Hence, the flux velocity can be treated as an offset to the constant
kinetochore velocity in the solution of the stationary FPEs.  The final effect
is a shift of both the master curves and the force-velocity relations by 
 $v_\mathrm{f}$ towards smaller kinetochore velocities $v_\mathrm{k}$
as shown in figure~\ref{fig:flux}(a). 
If the shift is so large
that the left turning point $F_\mathrm{min}$ of the force-velocity hysteresis
is located at a negative velocity,  poleward flux suppresses directional
instability because a fixed point emerges, and we expect 
similar behavior  as for intermediate
linker stiffnesses in the previous section (see figure\ \ref{fig:nonosc}).  In
the limit of many MTs, the maximum flux velocity that still allows directional
instability is given by the velocity in the maximum of the master curve, which
provides the boundary of the oscillatory regime in the parameter plane of
linker stiffness $c$ and poleward flux velocity $v_\mathrm{f}$
(figure~\ref{fig:flux}(b)).  Phase space diagrams (figure~\ref{fig:flux}(c)) and
kinetochore trajectories (figure~\ref{fig:flux}(d)) from simulations with
appropriate flux velocities confirm our arguments exhibiting similar 
behavior as for intermediate 
linker stiffnesses in  figure\ \ref{fig:nonosc}. 
For small flux velocities the boundary of the oscillatory 
regime in figure~\ref{fig:flux}(b) approaches our above result 
$c_\mathrm{osc}=\SI{15.91}{\pico\newton\per\micro\meter}$.
For increasing  flux velocities the oscillatory regime shrinks, and 
 its boundary has a maximum at
$c\approx\SI{50}{\pico\newton\per\micro\meter}$ with
$v_\mathrm{f}\approx\SI{3.11}{\nano\meter\per\second}$. We conclude  that
kinetochore oscillations can be suppressed by moderate flux
velocities independently of the linker stiffness.

Our theory also agrees with and explains  simulation results 
 in \cite{banigan2015}, 
where, for large flux velocities, 
suppression of kinetochore oscillations were observed 
but at the same time maintenance of
bistability. 
 Moreover, our results  explain the experimentally observed
correlation between flux velocity and directional instability.  Kinetochore
oscillations have been observed in the mitotic vertebrate cells listed in
table\ \ref{tab:flux} (LLC-PK1, PtK1/2, newt lung, U2OS) which  have 
poleward flux velocities not exceeding  $\SI{10}{\nano\meter\per\second}$,
whereas in the mitosis of a Drosophila embryo as well as in meiosis of a
Xenopus egg, where flux velocities are three to four times higher,
chromosomes do not exhibit directional instability.

\section{Polar ejection forces provide an alternating oscillation pattern
		and chromosome alignment at the spindle equator}
\label{sec:PEF}

So far,  we have not included polar ejection forces (PEFs). 
They originate from non-kinetochore MTs interacting with the
 chromosome arms and
pushing them thereby towards the spindle equator, either 
through collisions with the chromosome arms or via
chromokinesins~\cite{mazumdar2005}, and provide additional 
pushing forces on kinetochores. 
Therefore, they can be included into the model by 
adding forces $F_\mathrm{PEF,r}(X_\mathrm{k,r})$ and 
$F_\mathrm{PEF,l}(X_\mathrm{k,l})$ acting on kinetochores, 
which depend on the {\it absolute} position
of the kinetochores~\cite{civelekoglu2013}.
Due to the exponential length distribution of free MTs as well 
as the spherical geometry of the MT asters,
the density of non-kinetochore MTs decreases monotonically
 with the distance from the spindle pole.
Therefore, we assume that  PEFs reach their maximum at the centrosome 
and vanish at the spindle equator (located at $x=0$), where opposite PEFs 
compensate each other.
This assumption is supported by the monotonic PEF distribution
that has been measured in vivo by Ke {\it et al.}~\cite{ke2009}.
Here, we will only discuss linearized PEFs 
\begin{eqnarray}
	F_\mathrm{PEF,l}(X_\mathrm{k,l}) = -k X_\mathrm{k,l}, \qquad
	F_\mathrm{PEF,r}(X_\mathrm{k,r})&= k X_\mathrm{k,r},
\label{eqn:FPEF}
\end{eqnarray}
where the spring constant $k$ defines the strength of the forces,
and the signs are chosen so that a positive force acts in AP-direction.
We show in figure~S3 in the supplementary material
that other force distributions
do not differ qualitatively
in their influence on the kinetochore dynamics.

To determine kinetochore trajectories of the two-sided 
model in  the presence of PEFs,
we can start from  the same force-velocity relations 
 as for the basic one-sided model.
In the presence of PEFs,
the total forces $F_\mathrm{k,l}$ and $F_\mathrm{k,r}$ 
that act on the left and the right kinetochore in AP-direction
depend on the absolute kinetochore positions   $X_\mathrm{k,l}$ and 
 $X_\mathrm{k,r}$:
\begin{eqnarray}
	F_\mathrm{k,l} &= F_\mathrm{kk} ( \Delta X_\mathrm{k} ) + 
					F_\mathrm{PEF,l}(X_\mathrm{k,l}), \\
	F_\mathrm{k,r} &= F_\mathrm{kk} ( \Delta X_\mathrm{k} ) + 
					F_\mathrm{PEF,r}(X_\mathrm{k,r}).
\end{eqnarray}
We can investigate the motion of kinetochores in the full two-sided 
model again  by using a phase space diagram; in the presence of PEFs
we use a $v_\mathrm{k}$-$F_\mathrm{k}$-diagram 
with the total force $F_\mathrm{k}$ in AP-direction on the horizontal axis
and the velocity $v_\mathrm{k}$ in AP-direction on the vertical axis. 
Because the total forces contain the external PEFs they are 
no longer related by action and reaction and, thus, 
the two kinetochores no longer have the same position on the 
$F_\mathrm{k}$-axis, but they  still remain close to each other 
on the $F_\mathrm{k}$-axis as long as the cohesin bond is strong enough.

\begin{figure}[!h]
	\centering
	\includegraphics[width=\textwidth]{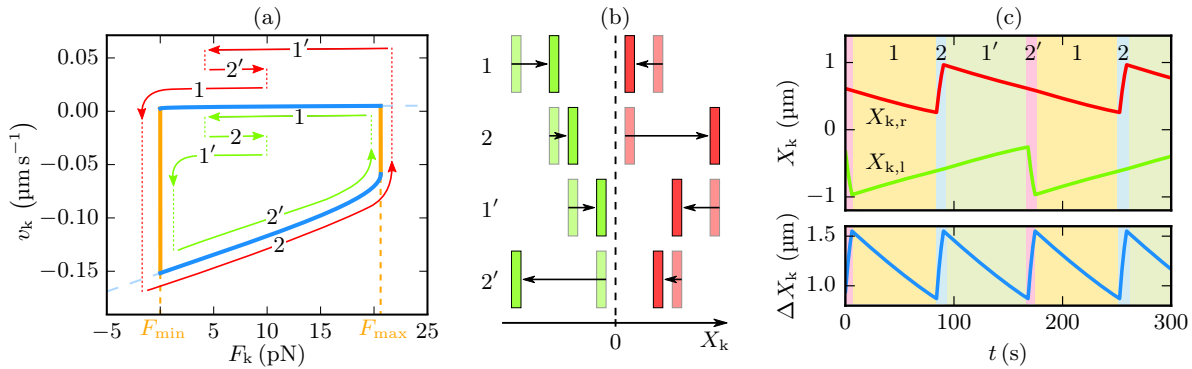}
	\caption{
            Kinetochore motion in the presence of PEFs.
            (a,b) At the beginning of state 1 the left kinetochore
            (green)  has just switched from P- to AP-movement,
			so that both kinetochores are on the upper branch.
			Both kinetochores move in AP-direction, which means
                        that both the cohesin force and the PEFs
			decrease and both kinetochores move left in the 
                       force-velocity diagram.
			Due to different PEFs, the right kinetochore (red) 
                      reaches the left turning point $F_\mathrm{min}$ first
			and switches to the lower branch, which marks the 
                       start of state 2.
			This state is dominated by the fast P-movement 
                 of the right kinetochore,
			which causes a steep increase of both
                         $F_\mathrm{kk}$ and $F_\mathrm{PEF,r}$.
			Therefore, the right kinetochore moves to the
                        right in the force-velocity diagram.
			Meanwhile, the left sister still moves in 
                  AP-direction and $F_\mathrm{k,l}$ increases slightly 
			as the increase of $F_\mathrm{kk}$ is larger 
                 than the decrease of $F_\mathrm{PEF,l}$.
			Since $\dot{F}_\mathrm{k,r} > \dot{F}_\mathrm{k,l}$,
                         the right kinetochore overtakes 
			its sister on the $F_\mathrm{k}$-axis before it
                       reaches the right turning point
			and switches to the upper branch.
			The then following states $1'$ and $2'$ are the 
                    exact opposite to 1 and 2 with swapped kinetochores.
			(c) Solution of the corresponding equations of 
                          motion for
			$c=\SI{20}{\pico\newton\per\micro\meter}$, 
                    $k=\SI{10}{\pico\newton\per\micro\meter}$ and $M=25$.
                    For an animated version see video~S4 in the supplementary material.
			}
	\label{fig:PEF}
\end{figure}

 A kinetochore on the upper/lower branch 
moves in AP-/P-direction with $v^\pm_\mathrm{k}(F_\mathrm{k})$ 
if $v^+_\mathrm{k}>0$ ($v^-_\mathrm{k}<0$). 
A kinetochore on the upper AP-directed branch will relax 
its AP-directed PEFs, while a kinetochore on the lower P-directed branch will 
build up AP-directed  PEFs.
After a time of equilibration the kinetochores behave as 
described in figure\ \ref{fig:PEF}.
When one kinetochore changes its direction from P to AP
(switches to the upper branch)
the sister kinetochore, which was on the upper branch before,
becomes the leading kinetochore
(here, ``leading'' refers to the position in the force velocity phase space). 
Therefore, the kinetochores do not reach the left turning point
$F_\mathrm{min}$  at the same time
so that it is always the leading kinetochore that 
switches to the lower branch.
Since in general the absolute  P-velocity is  much larger than 
the  AP-velocity ($-v_-$ for the lower branch is much larger than 
$+v_+$ for the upper branch),
the AP-directed PEF contribution to the 
total force  increases faster on the lower branch than on the upper one.
As a result, the P-moving kinetochore overtakes 
its sister on the $F_\mathrm{k}$-axis
before switching back to the upper branch such that 
the leading kinetochore automatically becomes the trailing 
kinetochore in the next oscillation period
(again, ``leading'' and ``trailing'' 
in terms of phase space positions).
This periodic 
change of kinetochore positions in the force-velocity diagram
leads to both regular breathing and regular single kinetochore oscillations,
as the kinetochores alternately pass the lower branch.
Solving appropriate equations of motions similar to \eref{eqn:eom} 
for each of the states depicted in figure\ \ref{fig:PEF}(ab),
we determine the deterministic 
trajectories in figure\ \ref{fig:PEF}(c) confirming this 
regular alternating oscillation pattern. 

\begin{figure}[!h]
	\centering
	\includegraphics{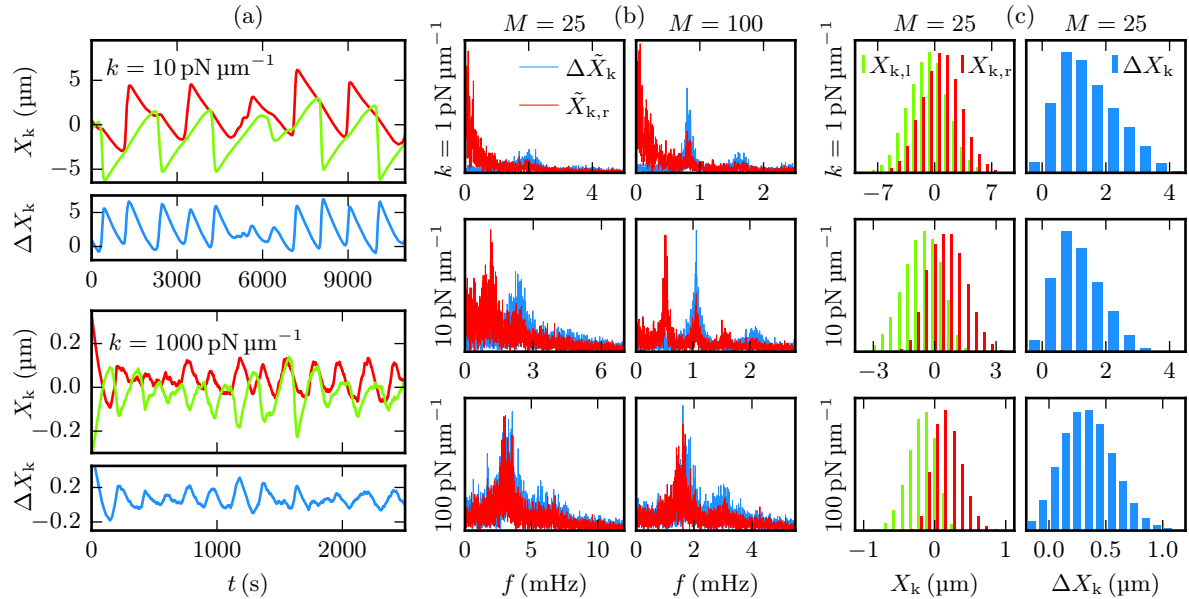}
	\caption{
	Kinetochore dynamics under the influence of PEFs.
	(a) Kinetochore trajectories with different PEF constants $k$
	from simulations with $M=100$, $c=\SI{20}{\pico\newton\per\micro\meter}$ 
	and without confinement at the spindle poles.
	The PEFs force the kinetochores to oscillate regularly
	and to stay near the spindle equator.
	For $k=\SI{10}{\pico\newton\per\micro\meter}$
	kinetochores oscillate as described in figure~\ref{fig:PEF}.
	Since with strong PEFs kinetochores tend to
	switch to the lower branch simultaneously
	when reaching $F_\mathrm{min}$ in the phase space at the same time,
	for $k=\SI{1000}{\pico\newton\per\micro\meter}$
	oscillations are in antiphase due to symmetric initial conditions
	before the system equilibrates at $t\approx\SI{1500}{s}$.
	After equilibration, periods of antiphase oscillations
	reappear over and over again due to fluctuations.
	Stronger PEFs cause a more fluctuative kinetochore motion.
	Especially for moderate MT numbers, 
	this can lead to suppression of kinetochore oscillations.
	For animated versions of phase space trajectories
	see videos~S5 ($k=\SI{10}{\pico\newton\per\micro\meter}$)
	and~S6 ($k=\SI{1000}{\pico\newton\per\micro\meter}$)
	in the supplementary material.
	(b) Single (right) kinetochore and breathing oscillations in Fourier space.
	For weak PEFs ($k=\SI{1}{\pico\newton\per\micro\meter}$)
	single kinetochore oscillations are still irregular
	and $\tilde{X}_\mathrm{k,r}$ has its maximum at $f=0$.
	If $k=\SI{10}{\pico\newton\per\micro\meter}$,
	$\tilde{X}_\mathrm{k,r}$ has a distinct peak at half the
	breathing frequency,
	indicating regular oscillations as described in figure~\ref{fig:PEF}
	and frequency doubling of breathing
	compared to single kinetochore oscillations.
	With sufficiently strong PEFs
	($k\gtrsim\SI{100}{\pico\newton\per\micro\meter}$)
	frequency doubling is lost as a consequence of antiphase oscillations
	and the peaks of $\tilde{X}_\mathrm{k,r}$ 
	and $\Delta\tilde{X}_\mathrm{k}$
	coincide with each other.
	(c) Histograms of kinetochore positions and inter-kinetochore distances
	for the realistic case of $M=25$.
	Chromosomes are aligned at the spindle
	equator despite missing confinement at the centrosome.
	The range of kinetochore positions is narrower
	and the distances smaller if PEFs are stronger.
	}
	\label{fig:PEFsimu}
\end{figure}

The alternating oscillation pattern robustly survives 
in stochastic simulations  
in the presence of moderate PEFs 
($k\sim\SI{10}{\pico\newton\per\micro\meter}$)
as we demonstrate in figure\ \ref{fig:PEFsimu}(a)
by means of the kinetochore trajectories in real space.
In figure~\ref{fig:PEFsimu}(b),
emergence of regular oscillations is illustrated
in Fourier space:
Whereas for rather small values of $k$
single kinetochore oscillations are still irregular
resulting in a nearly monotonic decreasing Fourier transform,
for $k=\SI{10}{\pico\newton\per\micro\meter}$
single kinetochore motion has a distinct peak in the Fourier space
indicating a regular shape of oscillations in real space.
Moreover, frequency doubling  of breathing
compared to single kinetochore oscillations
can directly be recognized
by comparing the corresponding Fourier transforms.
As a consequence of regular oscillations,
the kinetochores stay near the spindle equator
and can not get stuck to one of the centrosomes as in the basic model,
see histograms of kinetochore positions in figure~\ref{fig:PEFsimu}(c).
We conclude that PEFs are necessary to assure proper chromosome alignment
in the metaphase plate at the spindle equator.
This is consistent with an experiment by 
Levesque and Compton \cite{levesque2001},
who observed mitosis of vertebrate cells after suppressing 
the activity of chromokinesins and, thus PEFs.
This  resulted  in $\SI{17.5}{\percent}$ of the cells 
in at least one chromosome not aligning
at the equator, but locating near a spindle pole.

Moreover,  PEFs reduce 
the amplitude and increase the frequency of  oscillations.
The amplitude decreases for increasing PEF strength $k$ as the kinetochores 
have to cover a smaller distance
between the  turning points at  $F_\mathrm{min}$ and $F_\mathrm{max}$.
The increase of the frequency is linear in $k$,
which  can be deduced from the linear increase of $|\dot{F}_\mathrm{k}|$:
\begin{eqnarray}
	|\dot{F}_\mathrm{k,l}| &= 
   \left| c_\text{k} \left(v_\mathrm{k,r}+v_\mathrm{k,l}\right)
   	+ k v_\mathrm{k,l} \right|,\\
	|\dot{F}_\mathrm{k,r}| &= 
    \left| c_\text{k} \left(v_\mathrm{k,r}+v_\mathrm{k,l}\right)
	+ k v_\mathrm{k,r} \right|
\end{eqnarray}
(defining  $v_\mathrm{k,l}\equiv \dot{X}_\mathrm{k,l}$ and 
 $v_\mathrm{k,r}\equiv -\dot{X}_\mathrm{k,r}$ as the velocities
in AP-direction as before).

Since PEFs do not have any influence on the underlying master 
curves and force-velocity relations,
they do not affect the kinetochore velocities $v_\mathrm{k}$
and never completely suppress kinetochore oscillations
 in the deterministic Fokker-Planck model,
but only reduce their amplitude and increase their frequency.
For strong PEFs, however, this gives rise to 
 kinetochore motion with a fluctuative character,
 see figure\ \ref{fig:PEFsimu} (see also
 video~S6 in the supplementary material).
The same observation was made in the model of 
Civelekoglu-Scholey {\it et al.}~\cite{civelekoglu2013}.
Additionally, we detect sister kinetochore oscillations
being in antiphase if PEFs are strong enough
($k\gtrsim\SI{100}{\pico\newton\per\micro\meter}$),
see figure~\ref{fig:PEFsimu}(a).
This follows from the phase space velocities $\dot{F}_\mathrm{k}$
being dominated by the strong PEFs compared to inter-kinetochore tension:
Imagine, both kinetochores are in the upper branch of the phase space 
and reach the turning point $F_\mathrm{min}$ at nearly the same time.
When now one of the two kinetochores switches to the lower branch
and starts moving polewards,
its sister does not change its direction in phase space
as in state $2/2'$ in figure~\ref{fig:PEF}(a)
but continues moving left
since the decrease of PEFs due to its poleward motion
can not be compensated by the increasing AP-directed cohesin tension
if $k\gg c_\mathrm{k}$.
As a consequence, the kinetochore will switch to the lower branch
just after its sister and
both kinetochores pass the lower branch simultaneously,
i.e.\ move apart from each other,
finally resulting in antiphase oscillations.
While the antiphase behavior vanishes after a certain time of equilibration
in the deterministic model,
in stochastic simulations 
periods of antiphase oscillations can be observed over and over again
regardless of whether the system has been equilibrated before.
A characteristic of antiphase oscillations is the loss of frequency doubling
which also appears in the Fourier space where
the peaks of single kinetochore and breathing motion coincide with each other
if PEFs are strong,
see figure~\ref{fig:PEFsimu}(b).
Since antiphase kinetochore oscillations have not been observed experimentally,
we conclude that in vivo PEFs are weak
compared to the inter-kinetochore tension
but strong enough to assure chromosome alignment at the spindle equator.
Compared to experimental
results~\cite{skibbens1993,waters1996,magidson2011,wan2012,dumont2012,civelekoglu2013},
in our model, $k=\SI{10}{\pico\newton\per\micro\meter}$
seems a reasonable choice
as it assures regular oscillations with frequency doubling,
keeps the inter-kinetochore distance within a suitable
range of $\SI{1.2+-0.7}{\micro\meter}$,
and aligns kinetochores in a realistic maximum distance of
$\SI{3}{\micro\meter}$ from the spindle equator
with a standard deviation of $\SI{0.88}{\micro\meter}$
in the lifelike case of $M=25$.

\section{Catastrophe promotion at the kinetochore is required to stimulate
  directional instability if microtubules can not exert pushing forces}
\label{sec:push}

So far, we assumed that  MTs  are also able to exert pushing forces
on  the kinetochore. During oscillations 
we find, on average, slightly less (48\%)
 MT-kinetochore links under tension, while
a substantial part of linkers also exerts pushing forces.
Two experimental results suggest, however, that 
MTs do not directly exert pushing forces on the kinetochore:
In \cite{waters1996}, it was shown that the link between
chromosomes is always under tension;
the experiments in \cite{khodjakov1996} demonstrated
that, after removal of the cohesin bond,  AP-moving kinetochores
 immediately stop indicating that kinetochore MTs can not exert pushing forces, 
while P-moving kinetochores continue moving due to 
MT pulling forces.

In view of these experimental results
and in order to answer the question whether MT pushing forces are 
essential for bistability and oscillations,
we analyze variants of our basic model, where  
MT  growth is confined at the kinetochore, i.e., 
where the relative coordinate $x= x_\mathrm{m}-X_\mathrm{k}$ is 
limited to $x\le 0$ such that MTs can only exert tensile 
forces on the kinetochore. 
This implies that the kinetochore undergoes a catastrophe 
if it reaches the kinetochore, i.e., if the relative 
coordinate reaches $x=0$ from below in the one-sided model.
Different choices for the corresponding catastrophe rate 
$\omega_\mathrm{c}^\mathrm{kin}$ at $x=0$ are 
possible:
(i) A reflecting boundary, i.e., 
  $\omega_\mathrm{c}^\mathrm{kin}= \infty$, where a catastrophe 
is immediately triggered if the MT plus-end reaches the 
kinetochore.
(ii) A ``waiting'' boundary condition, where the {\it relative} velocity 
$v_+ = v_{\mathrm{m}+}-v_\mathrm{k}=0$ stalls if the MT reaches 
$x=0$
(in the simulation, 
 we set the MT velocity to $v_{\mathrm{m}+} = v_\mathrm{k}$).
In contrast to the reflecting boundary condition,
 the catastrophe rate $\omega_\mathrm{c}^\mathrm{kin}$ at the 
kinetochore is finite such that the MT waits at the kinetochore 
until it  undergoes a catastrophe
for  a mean  waiting time $1/\omega_\mathrm{c}^\mathrm{kin}$,
as similarly observed in metaphase of PtK1 cells~\cite{vandenbeldt2006}.
Because $x=0$ also results in $F_\mathrm{mk}=0$, the 
force-free catastrophe rate seems a natural choice, 
  $\omega_\mathrm{c}^\mathrm{kin}= \omega^0_\mathrm{c}$ 
[see (\ref{eq:expF})], which should be realized 
in the absence of any additional catastrophe regulating proteins
at the centromere. 
(iii)
If catastrophes are promoted by regulating proteins, but
not immediately as for (i), we obtain intermediate cases 
of waiting boundary conditions 
with $\omega^0_\mathrm{c} < \omega_\mathrm{c}^\mathrm{kin}< \infty$.
In mammalian cells, such 
regulating  mechanisms could be provided by the kinesin MCAK,
which is localized at the centromere during metaphase~\cite{wordeman1995}
and has been reported to  increase the catastrophe rate of MTs
roughly 7-fold \cite{newton2004}.
Therefore, waiting boundary 
conditions with an increased catastrophe rate appear to be the most 
realistic scenario. 
We introduce a numerical catastrophe enhancement 
factor $n\ge 1$ characterizing the increased catastrophe rate, 
 $\omega_\mathrm{c}^\mathrm{kin}=n\omega^0_\mathrm{c}$.
Within this general scenario 
reflecting boundary conditions (i) are recovered 
for  $n=\infty$ and (ii) waiting boundary conditions with 
the zero force catastrophe rate for $n=1$. 
We will discuss the general case (iii) in the following. 

In our basic model, where MTs can exert pushing forces on kinetochores,
the  pushing phases where $x>0$  can also be interpreted as a 
an effective waiting phase at the kinetochore with a catastrophe 
rate, which is effectively increased by the pushing forces. 
Therefore, the behavior of 
our basic model resembles a model with waiting boundary 
conditions with an increased catastrophe rate $n>1$ at the 
kinetochore. 
MT pushing forces are not 
essential for bistability and oscillations and have a 
similar effect as an increased catastrophe rate at the 
kinetochore as our detailed analysis will show.

In the Fokker-Planck solution for the one-sided model,
all confining  boundary conditions limit
the maximum MT-kinetochore distance  $x_\mathrm{max}$ to zero,
where  it is positive in the basic model.
When  $x_\mathrm{max}$ is negative in the basic model
(for $v_\mathrm{k} > v^0_+$, see table~\ref{tab:boundary}),
confining boundary conditions 
 do not modify the basic model,
since the MTs are not able to reach the fast kinetochore.
For negative kinetochore velocities  $v_\mathrm{k} <v^0_-$,
the minimum distance $x_\mathrm{min}$ becomes positive
while $x_\mathrm{max}$ is zero.
Then, all confining   boundary conditions 
fix the  MT tips to the kinetochore position 
as they do not shrink fast enough to move away 
from the poleward-moving kinetochore after a catastrophe
resulting in $\langle x \rangle=0$ and $F_\mathrm{ext} = \gamma v_\mathrm{k}$.
All in  all, confinement leads to the following maximal and minimal values
 for the MT-kinetochore distance $x$ modifying table~\ref{tab:boundary}:
\begin{eqnarray}
	x^\mathrm{conf}_\mathrm{max} =
		\cases{
			0, & $v_\mathrm{k}<v^0_+$\\
			x_\mathrm{max}, & $v_\mathrm{k}\geq v^0_+$,}
	\qquad
	x^\mathrm{conf}_\mathrm{min} = 
		\cases{
			0, & $v_\mathrm{k}<v^0_-$\\
			x_\mathrm{min}, & $v_\mathrm{k}\geq v^0_-$.
		}
\end{eqnarray}

\begin{figure}[!h]
	\centering
	\includegraphics{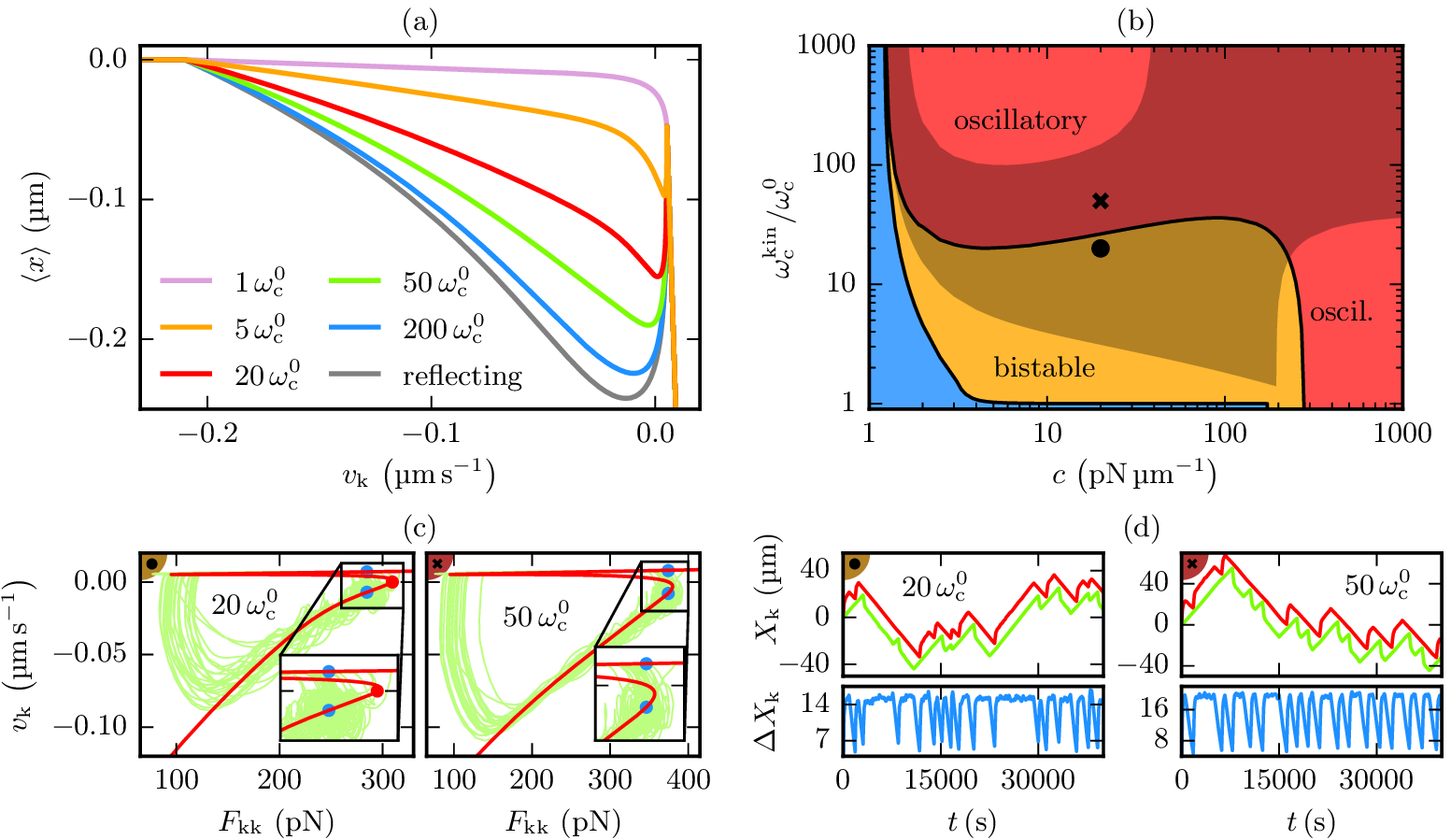}
	\caption{
		Microtubule confinement at the kinetochore.
		(a) Master curves of a system with a waiting boundary
		condition for various $\omega_\mathrm{c}^\mathrm{kin} =
		n\,\omega_\mathrm{c}^0$ and
		$c=\SI{20}{\pico\newton\per\micro\meter}$.
		(b)~Regimes in the parameter plane of $c$ and 
      $\omega_\mathrm{c}^\mathrm{kin}$
		in the limit of many MTs.
		Outside the blue region, the master curve is bistable.
		In the orange region, the left branch of the master curve 
		and, therefore, the lower branch of the 
		$v_\mathrm{k}$-$F_\mathrm{kk}$-diagram cross $v_\mathrm{k}=0$,
		which leads to a fixed point  suppressing oscillations 
         (see text),
		whereas in the red region oscillations are possible.
		In stochastic simulations, kinetochores already oscillate
		at much smaller $\omega_\mathrm{c}^\mathrm{kin}$
		than predicted by the master curves.
		Additionally, a new kind of fixed point, 
               which is depicted in (c),
		emerges in the shaded region.
		(c,d) Phase space diagrams and kinetochore trajectories
		from simulations of the unconfined two-sided model with
		$c=\SI{20}{\pico\newton\per\micro\meter}$ and $M=100$. 
		The blue dots mark the new kind of fixed point,
		where the leading kinetochore in the
		lower branch moves with the same velocity as the trailing
		kinetochore in the upper branch.
		Then the inter-kinetochore distance remains constant,
		while the center of mass moves with a 
              constant velocity as in (d) for 
		$\omega_\mathrm{c}^\mathrm{kin} = 20\,\omega_\mathrm{c}^0$ 
		at $t\approx\SI{25000}{s}$.
		In the presence of PEFs, these fixed points are absent
		and the shaded region in (b) does not apply.
	}
	\label{fig:conf_wckin}
\end{figure}

We  calculate the master curves 
$\langle x\rangle (v_\mathrm{k})$ for all three types of confining 
boundary conditions (see figure\ \ref{fig:conf_wckin}(a)). 
Because $x^\mathrm{conf}_\mathrm{max}\le 0$ for any confining 
boundary condition, also 
$\langle x \rangle < 0$, i.e., the complete master curves 
lie in the regime of tensile MT-kinetochore linker  forces
reflecting the fact that 
 pushing forces are strictly 
suppressed. 
Therefore, the MT-kinetochore catch 
bond is on average  under tension 
 establishing a more firm MT-kinetochore connection 
 during the stochastic chromosome oscillations in metaphase.
Oscillations then become a tug-of-war, in which both sets of MTs 
only exert  pulling forces onto each other.  

With a waiting boundary condition at the kinetochore,
the probability densities $p_\pm(x,t)$ have to be supplemented
with the probability $Q(t)$ to find a MT at the kinetochore ($x=0$).
Besides the FPEs \eref{eqn:FPG+} and \eref{eqn:FPG-} 
for the probability densities,
we also have to solve the equation for the time evolution of $Q(t)$:
\begin{eqnarray}
	\partial_t Q(t) = v_+(0) p_+(0,t) - 
  \omega_\mathrm{c}^\mathrm{kin}Q(t). \label{eqn:dtQ}
\end{eqnarray}
The analogous model for a free MT that grows against a rigid wall
has already been solved in \cite{mulder2012,Zelinski2012}.
In the stationary state, \eref{eqn:dtQ} leads to
$Q =  p_+(0){v_+(0)}/{\omega_\mathrm{c}^\mathrm{kin}}$.
For the probability densities $p_\pm(x)$ we get the same solution
as for the basic model without confinement, 
except for the normalization constant.
The overall probability density can then be written as
$p(x) = p_+(x) + p_-(x) + Q\delta(x)$ and has to satisfy
$\int_{x^\mathrm{conf}_\mathrm{min}}^{x^\mathrm{conf}_\mathrm{max}} p(x) \dd x = 1$.

From the overall probability density 
$p(x)$ we obtain the master curves, which we show  
in figure\  \ref{fig:conf_wckin}(a) 
 for $n=1,5,20,50,200,\infty$ 
and  a linker stiffness of $c=\SI{20}{\pico\newton\per\micro\meter}$.
Again we  can analyze the master curves for extrema to
obtain constraints on  linker stiffness  $c$ and  catastrophe
enhancement factor $n = \omega_\mathrm{c}^\mathrm{kin}/\omega^0_\mathrm{c}$
for the occurrence of bistability and oscillations.
The results of this analysis 
are shown in figure~\ref{fig:conf_wckin}(b) as colored regions.
It turns out that extrema in the master curve and, thus, bistability 
occur if  the linker stiffness is sufficiently high $c>c_\mathrm{bist}$. 
For the zero force catastrophe rate  $n=1$  we find a high threshold 
value
 $c_\mathrm{bist} = \SI{178}{\pico\newton\per\micro\meter}$,
in the limit of a reflecting boundary $n=\infty$ a very low
threshold  $c_\mathrm{bist} = \SI{1.218}{\pico\newton\per\micro\meter}$.

 We remind that  a sufficient condition for oscillations 
is the absence of a stable fixed point, where  one 
 of the two branches in the $v_\mathrm{k}$-$F_\mathrm{kk}$-diagram 
crosses $v_\mathrm{k}=0$.
In contrast to the basic model, 
the maxima of the master curve are now located at a positive velocity
for $n>1$.
Therefore, oscillations are suppressed by a fixed point 
 $v^-_\mathrm{k}=0$ on the lower branch in 
the $v_\mathrm{k}$-$F_\mathrm{kk}$-diagram,
 which occurs if the velocity is positive in the 
 minimum of the master curve.
In general, oscillations occur 
 if  the linker stiffness is sufficiently high $c>c_\mathrm{osc}$. 
Again we find a high threshold value 
 $c_\mathrm{osc} = \SI{280}{\pico\newton\per\micro\meter}$
for $n=1$ and a low threshold  
$c_\mathrm{osc} = \SI{1.237}{\pico\newton\per\micro\meter}$
for a reflecting boundary 
condition ($n=\infty$).

For $n<10$ the threshold values remain high. 
Moreover, 
at such high linker
stiffnesses and for for small $n$, the 
 simulations of the two-sided model do not show the expected behavior.
For $n=1$ and 
 high linker stiffnesses in the oscillatory regime 
 the kinetochore trajectories do not exhibit regular oscillations.
 Naively, one could argue that kinetochore oscillations are suppressed
due to the lack of a pushing force and can be restored by additional PEFs.
However, this is not the case, since, as stated above,
PEFs do not affect the master curve that 
determines the regime of kinetochore motion.
 One reason for the absence of oscillations is that, for the zero force
catastrophe rate ($n=1$) the 
waiting time $1/\omega_\mathrm{c}^\mathrm{kin}\sim 500{\rm s}$ (see 
 table\ \ref{tab:parameters}) at the kinetochore is  large compared
 to the typical oscillation periods, which are in the range of 
 $100-200{\rm s}$.

Figure~\ref{fig:conf_wckin}(b) also shows that 
 oscillations require increased catastrophe rates 
with  $n\gtrsim 20$ over a wide range of linker stiffnesses 
from $c=\SI{10}{\pico\newton\per\micro\meter}$ to 
$c=\SI{200}{\pico\newton\per\micro\meter}$. 
For $n>1$, at the boundary between bistable and oscillatory regime 
 in figure~\ref{fig:conf_wckin}(b),  a fixed point 
$v^-_\mathrm{k}=0$ on the lower branch of the $v_\mathrm{k}$-$F_\mathrm{kk}$ 
 phase space diagrams appears, which can suppress oscillations.  
This fixed point is, however, less relevant
because the kinetochores will only occasionally 
pass the lower branch simultaneously,
which is necessary to  reach this fixed point.
Furthermore, this fixed point is located 
near the right turning point $F_\mathrm{max}$
so that the kinetochores can easily leave the fixed point
by a stochastic fluctuation (as in figure~\ref{fig:nonosc}).
For these two reasons, in stochastic simulations,
oscillations already occur for $n\gtrsim 5$, that is 
at a much lower $n$ than the deterministically predicted $n\gtrsim 20$,
 but  not for $n=1$, i.e., in the absence of a catastrophe promoting 
mechanism.

The fixed point analysis of the $v_\mathrm{k}$-$F_\mathrm{kk}$ 
 phase space diagrams
reveals that also   a new type of fixed point corresponding to a 
non-oscillatory motion  emerges for 
$n\lesssim 100$ in the shaded regions in  figure~\ref{fig:conf_wckin}(b). 
In this new type of 
fixed point,  the leading P-moving kinetochore in the
 lower branch of the master curve 
  has the same velocity as the trailing
 AP-moving  kinetochore in the upper branch  
(see figure\ \ref{fig:conf_wckin}(c))
 so that  $\dot{F}_\mathrm{kk} = -c_\text{k}
 \left(v_\mathrm{k,r}+v_\mathrm{k,l}\right)=0$, and 
   the inter-kinetochore
    distance remains constant, while
    the center of mass moves with a constant velocity 
  (see figure\ \ref{fig:conf_wckin}(d)). 
In the presence of PEFs, however,   this new type of fixed point 
does not survive because for the  P- moving kinetochore 
the AP-directed   PEFs increase, whereas they decrease 
for an AP-moving kinetochore. Then 
the upper blue dot in  figure\ \ref{fig:conf_wckin}(c) moves to the 
left, while the lower blue point moves to the right such that this
new type of fixed point is unstable in the presence of PEFs.
Therefore,  in the entire shaded region
 in   figure~\ref{fig:conf_wckin}(b)  PEFs  are essential to 
re-establish oscillations.

We conclude that  both the linker stiffness 
$c>\SI{10}{\pico\newton\per\micro\meter}$ and
the catastrophe rate $\omega_\mathrm{c}^\mathrm{kin}$ at the kinetochore 
($n\gtrsim 20$ or $n\gtrsim 5$ in the presence of stochastic fluctuations)
have to be sufficiently large to obtain bistability and oscillations.
Because additional catastrophe promoting proteins are necessary 
to increase the catastrophe rate at the kinetochore, the lowest values
of $n$, which still enable oscillations, 
 might be  advantageous in the cellular system. 
   We note that poleward flux can
   influence existence and positions of fixed points:
An intermediate flow velocity
 can eliminate a fixed point on the lower branch 
by moving it into the unstable area of the phase space diagram.
If flux is sufficiently large 
it can establish additional fixed points on the upper branch 
of the phase space diagrams, which suppress oscillations 
as in the basic model.

Moreover, the linker stiffness has  to be sufficiently high 
to give  linker extensions compatible with experimental results.
 An important part of the MT-kinetochore linkage is Ndc80,
 which is a rod-like fibril of total length 
 around $60\,{\rm nm}$ \cite{Wei2005,Wang2008}
  consisting of two coiled-coil regions with a flexible hinge
 that  can adopt bending angles up to $120^\circ$ with a broad distribution
 \cite{Wang2008}. This bending corresponds to linker length changes 
  of $|x|\sim \SI{50}{\nano\meter}$. Moreover, 
 fluorescent labeling showed  total  intra-kinetochore stretches 
 around  $\SI{100}{\nano\metre}$ \cite{Maresca2009}  or
 $\SI{50}{\nano\metre}$ \cite{dumont2012}.
 Therefore, we regard  linker 
extensions $x\lesssim \SI{100}{\nano\meter}$ as realistic values. 
For large $n\gg 20$  only a small linker stiffness is 
necessary to enable oscillations. 
At the small threshold  stiffness, the average linker length 
 $|\langle x \rangle|$ 
 is  typically $\SI{1}{\micro\meter}$ in this regime.  
Increasing the linker stiffness leads to a decreasing 
 linker length $|\langle x \rangle|$. 
 We conclude that, for $n\gg 20$, 
 experimental observations of linker extensions 
 $|x|\lesssim \SI{100}{\nano\meter}$ put a stronger constraint 
 on  linker stiffness  than the experimental observations of oscillations. 
 Linker stiffnesses significantly above 
 $\SI{5}{\pico\newton\per\micro\meter}$ and, thus, far above $c_\mathrm{osc}$
 are necessary to obtain a realistic linker length.

For $n\sim 10-20$, which is 
compatible with the experimental result $n\sim 7$ for the catastrophe promoter
MCAK \cite{newton2004}, and a linker stiffness 
  $c=\SI{20}{\pico\newton\per\micro\meter}$,
the increased catastrophe rate at the kinetochore 
leads to a realistic behavior  with  linker 
extensions $x\sim \SI{100}{\nano\meter}$, which are also  compatible with 
the experimental results \cite{Wei2005,Wang2008,Maresca2009,dumont2012}
(see  figure\ \ref{fig:conf_wckin}(a)).
This parameter regime is within the shaded regions 
 in   figure~\ref{fig:conf_wckin}(b) and 
PEFs are necessary to establish oscillations.
The linker extension is independent of PEFs. 

 For an increased catastrophe rate around $n\sim 10-20$
 and a linker stiffness  $c=\SI{20}{\pico\newton\per\micro\meter}$,
 the more realistic model with waiting boundary conditions 
at the kinetochore exhibits a similar behavior as our 
basic model
because pushing phases  where $x>0$ in the basic model 
 have a similar duration 
as waiting times at the kinetochore in the more realistic model.

\section{Model parameters can be adjusted to 
	reproduce kinetochore oscillations in PtK1 cells}
\label{sec:ptk1}

So far,
we took the experimentally measured parameters
for MT transitions and velocities from table~\ref{tab:parameters}
for granted
in order to analyze the effects of poleward flux,
PEFs and confinement at the kinetochore
by means of our mean-field theory.
These values stem from experiments with yeast kinetochores~\cite{akiyoshi2010},
which can only bind one MT~\cite{winey1995},
whereas the mean-field theory is only correct
if the kinetochores are attached to multiple MTs
as in metazoan cells.
Moreover, in budding yeast,
the Ndc80 fibrils are connected to MTs 
via ring-like Dam1 complexes, which do not appear in
metazoan cells~\cite{cheeseman2014}.
In this  section, 
we  demonstrate
that by adjusting the parameters of MT dynamics
our model can reproduce 
 experimental data of metazoan spindles
using the example of PtK1 cells.

Our model exhibits a 
 large difference of P versus AP-velocity
($\sim100$ vs.\ $\sim\SI{4}{\nano\meter\per\second}$,
see figure~\ref{fig:phasediagram_osc})
which is the origin of frequency doubling and also
appears in PtK1 cells but not in this extent
($\sim19$ vs.\ $\sim\SI{16}{\nano\meter\per\second}$)~\cite{wan2012}.
As a consequence, in our model
both kinetochores move towards each other in AP-direction
(state 0 in figure~\ref{fig:osc}) most of the time,
whereas in the experiment, mostly one kinetochore moves in P-
while the trailing sister is moving in AP-direction
(state $2/2'$ in figure~\ref{fig:osc}).
In a first step we will respect these results
by adjusting the master curve (or force velocity relation)
in a way that the two stable
branches fit the experimentally measured velocities.
This objective will be achieved 
by modifying the force-free MT velocities $v_\pm^0$
(shifting the upper / lower branch up- or downwards)
and the corresponding characteristic forces $F_\pm$
(altering the slope of the upper / lower branch).
Moreover, as a last parameter of MT dynamics,
we will change the rescue rate $\omega_\mathrm{r}^0$
in order to adjust the MT-kinetochore distance to a realistic value.
In a second step we will fit the measured frequencies
and amplitudes by varying the parameters that do not
affect the master curves ($c_\mathrm{k}$, $k$).

Using the model with confinement at the kinetochore,
we assume a ten times increased catastrophe rate 
$\omega_\mathrm{c}^\mathrm{kin}=10\omega_\mathrm{c}^0$
according to experimental results~\cite{newton2004}.
We set the linker stiffness to $c=\SI{20}{\pico\newton\per\micro\meter}$
and keep it unchanged henceforth
since this value results in strongly bistable master curves
and the manifold consequences
that a further modification of $c$ has on kinetochore dynamics
are hard to handle.
The flux velocity is $v_\mathrm{f}=\SI{8}{\nano\meter\per\second}$
(see table~\ref{tab:flux}).
The force-free MT growth velocity $v_+^0$ has to be greater
than $v_\mathrm{f}$ for two reasons:
Firstly, detached MTs would not have a chance to reach the kinetochore again,
otherwise.
Secondly, this choice prevents a fixed point at the upper branch,
as the left turning point in phase space (maximum of the master curve)
is located at $v_+^0 - v_\mathrm{f}$,
when the MTs are confined at the kinetochore.
We increase the force-free growth velocity roughly four-fold
to $v_+^0=\SI{20}{\nano\meter\per\second}$,
so that the minimum AP-velocity
$v_+^0 -v_\mathrm{f} = \SI{12}{\nano\meter\per\second}$
in the left turning point $F_\mathrm{min}$
lies below the observed mean velocity of 
$\sim\SI{16}{\nano\meter\per\second}$.
In order to adjust the maximum AP-velocity,
we reduce the characteristic force in MT growth
to $F_+=\SI{5}{pN}$,
which leads to a steeper upper branch in the  phase space diagram.
The force-free shrinking velocity $v_-^0$
should be smaller than the observed P-velocity
since the lower, P-directed branch always lies above it.
Analogously to the upper branch and $F_+$,
also the slope of the lower branch can be adjusted
by varying the characteristic force $F_-$:
An increase of $F_-$, i.e.\ a decrease of its absolute value,
steepens the lower branch and thereby 
slows down the poleward motion.
It turns out that it is a good choice to keep the values
for $v_-^0$ and $F_-$ from table~\ref{tab:parameters}
unchanged.
Finally, we reduce the rescue rate $\omega_\mathrm{r}^0$,
which lets MTs shrink to smaller lengths  $x_\mathrm{m}$
(the minimum of the master curve is shifted downwards)
and increases the
MT-kinetochore distance $|x| = |X_\mathrm{k}-x_\mathrm{m}|$
to a realistic value.

Since we enable detachment in this section,
we set $M=35$ as it results in
a mean number of $\sim20$ attached MTs.
Finally, we adjust the strength of PEFs $k$ and
the cohesin bond stiffness $c_\mathrm{k}$
to the following conditions:
Firstly, the PEFs have to be strong enough to
assure proper chromosome alignment at the equator
as well as a regular oscillation pattern,
but should not dominate compared to the inter-kinetochore tension
in order to prevent antiphase oscillations.
Secondly, $k$ and $c_\mathrm{k}$ affect the amplitude
and the frequency of kinetochore oscillations
which should resemble experimental results
in the same manner:
An increase of both $k$ and $c_\mathrm{k}$
decreases the amplitude and increases the frequency.
We find that $k=\SI{20}{\pico\newton\per\micro\meter}$ and
$c_\mathrm{k}=\SI{20}{\pico\newton\per\micro\meter}$
 fulfill both conditions.
In table~\ref{tab:ptk1parameters}, we list all
 parameters that we have changed 
compared to table~\ref{tab:parameters}.

\begin{table}[!ht]
	\centering
	\caption{
	Parameters to  reproduce of kinetochore
	oscillations in PtK1 cells.
	Parameters  not listed here
	have been unchanged compared to table~\ref{tab:parameters}.}
	\label{tab:ptk1parameters}
	\begin{indented}
	\lineup
	\item[]\begin{tabular}{lll}
		\br
		Description & Symbol & Value \\
		\mr
		zero force rescue rate & $\omega_\mathrm{r}^0$ & \SI{0.012}{\per\second} \\
		zero force MT growth velocity & $v_+^0$ & \SI{20}{\nano\meter\per\second} \\
		characteristic force of MT growth  & $F_+$ & \SI{5}{pN}\\
		catastrophe rate at the kinetochore & $\omega_\mathrm{c}^\mathrm{kin}$ & \SI{0.019}{\per\second} \\
		MT flux velocity & $v_\mathrm{f}$ & \SI{8}{\nano\meter\per\second} \\
		PEF coefficient & $k$ & \SI{20}{\pico\newton\per\micro\meter} \\
		cohesin bond stiffness & $c_\mathrm{k}$ & \SI{20}{\pico\newton\per\micro\meter} \\
		MT-kinetochore linker stiffness & $c$ & \SI{20}{\pico\newton\per\micro\meter} \\
		number of MTs & $M$ & 35\\
		\br
	\end{tabular}
	\end{indented}
\end{table}

\begin{figure}[t]
	\centering
	\includegraphics{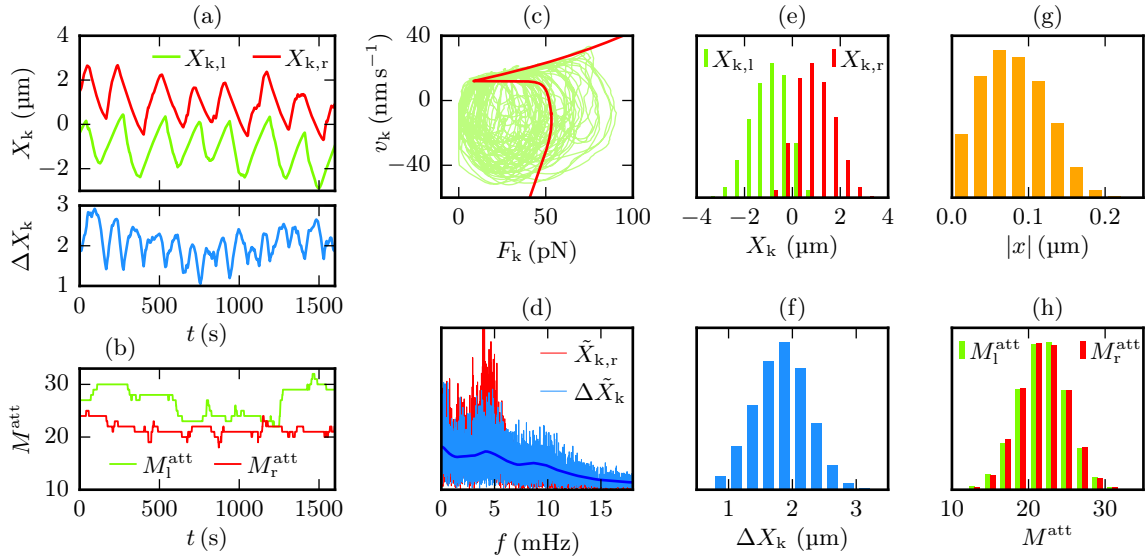}
	\caption{
		Reproduction of kinetochore oscillations in PtK1 cells.
		(a) Kinetochore positions and inter-kinetochore distance over time.
		Although the breathing oscillations are rather fluctuative,
		frequency doubling is recognizable.
		(b) Number of attached MTs over time.
		(c) Kinetochore motion in phase space (green)
		compared to the mean-field force-velocity relation (red,
                calculated with the 
		mean number of attached MTs).
                For an animated version see video~S7 in the supplementary material.
		(d) Position of the right kinetochore and inter-kinetochore distance
		in Fourier space.
		Fluctuative breathing oscillations lead to a 
		Fourier transform with broad maxima,
		which are almost only recognizable in the smoothed curve (dark blue).
		(e-h) Distributions of kinetochore positions $X_\mathrm{k}$,
		inter-kinetochore distance $\Delta X_\mathrm{k}$,
		MT-kinetochore distance $|x|$,
		and the number of attached MTs $M^\mathrm{att}$.
		}
	\label{fig:ptk1}
\end{figure}

The resulting kinetochore dynamics is shown in figure~\ref{fig:ptk1}.
The simulated kinetochore trajectories
in figure~\ref{fig:ptk1}(a) are very similar
to the experimental results in\ \cite{wan2012,civelekoglu2013}
as they exhibit frequency doubling of breathing compared
to single kinetochore oscillations
and move predominantly in phase,
i.e.\ there is a leading P- and a trailing AP-kinetochore
(state $2/2'$ in figure~\ref{fig:osc}).
The motion of the inter-kinetochore distance is rather fluctuative,
resulting in a broad Fourier transform,
in which the maximum at the breathing frequency is hardly recognizable,
see figure~\ref{fig:ptk1}(d).
This is the only significant difference to the real kinetochore motion.
The distributions of kinetochore positions as well as 
inter-kinetochore and MT-kinetochore distances (figure~\ref{fig:ptk1}(e-g))
are in good agreement
with experimental results~\cite{civelekoglu2013}.

In table~\ref{tab:ptk1comparison},
we list several characteristic quantities of kinetochore oscillations
that have also been determined experimentally for PtK1 cells.
Comparison with our model results shows quantitative agreement.
In particular, the large discrepancy in the P- and AP-velocities
is eliminated.

\begin{table}[!ht]
	\centering
	\caption{
	Characteristic quantities of model kinetochore oscillations
	compared to experimental results in PtK1 cells.}
	\label{tab:ptk1comparison}
	\begin{indented}
	\lineup
	\item[]\begin{tabular}{llll}
		\br
		Description & Model & Experiment& \\
		\mr
		mean P velocity & \SI{21.5}{\nano\meter\per\second} & \SI{19.0}{\nano\meter\per\second} 
			& \cite{wan2012} \\
		mean AP velocity & \SI{15.7}{\nano\meter\per\second} & \SI{15.7}{\nano\meter\per\second} 
			& \cite{wan2012} \\
		single kinetochore frequency & \SI{4.27}{\milli\hertz} & 4.14--\SI{4.23}{\milli\hertz} & \cite{wan2012} \\
		breathing frequency & $\sim\SI{8.6}{\milli\hertz}$ & \SI{8.25}{\milli\hertz} & \cite{wan2012} \\
		mean inter-kinetochore distance & \SI{1.83+-0.42}{\micro\meter} & \SI{1.90+-0.44}{\micro\meter} 
			& \cite{civelekoglu2013} \\
		mean MT-kinetochore distance & \SI{0.081+-0.042}{\micro\meter} & \SI{0.11+-0.04}{\micro\meter} 
			& \cite{civelekoglu2013} \\
		standard deviation of kinetochore position & 
			\SI{0.76}{\micro\meter} & 0.5--\SI{1.1}{\micro\meter} & \cite{civelekoglu2013} \\
		mean number of attached MTs & 21.4 & 20--25 & \cite{mcewen1997} \\
		\br
	\end{tabular}
	\end{indented}
\end{table}

\section{Discussion}

We provided an analytical mean-field solution  of 
the one-sided spindle model introduced by  
Banigan {\it et al.}~\cite{banigan2015}, 
which becomes  exact
 in the limit of large MT numbers. 
The mean-field solution is based on the calculation of 
the  mean linker extension $\langle x \rangle$ 
as a function of a constant kinetochore velocity 
$v_\mathrm{k}$ (the master curve). 
Together with the equation of motion of the kinetochore we obtained 
the force-velocity relation of the one-sided model from the master curve. 
  Our solution clearly shows that the force feedback
   of linkers onto the MT depolymerization dynamics is essential for a
   bistable force-velocity relation within the minimal model.
   The shape of the distribution $p_\pm(x)$ of linker lengths  (\ref{eqn:p+-})
   is governed by this force feedback, and we traced
   the bistability to the peakedness (kurtosis)
    of this distribution. 

Bistability of the force-velocity relation in the one-sided model
is a necessary (but not sufficient) 
condition  for oscillations in the two-sided model.
Interpreting the
bistable force-velocity relation as phase space diagram, we 
mathematically described kinetochore oscillations as an emergent result
of collective dynamics of coupled MTs that exhibit 
dynamic instability individually. Our theory becomes exact in the limit
of large MT numbers $M$.
This interpretation of oscillations
is underpinned by the experimental observations
that kinetochore oscillations in
budding yeast~\cite{straight1997,pearson2001,sprague2003},
where each kinetochore is attached to one MT~\cite{winey1995},
as well as in fission yeast~\cite{nabeshima1998,klemm2018}, 
where two to four MTs interact with the same kinetochore~\cite{ding1993},
have a considerably more fluctuative character
than the regular oscillations in vertebrate
cells~\cite{skibbens1993,waters1996,mitchison1989,ganem2005,magidson2011,wan2012,dumont2012}
with $\sim 20$ MTs per kinetochore~\cite{mcewen1997,mcewen2001}.
Moreover,
we were able to deduce idealized kinetochore oscillations,
whose periods conform with
experimental results~\cite{wan2012}.   For a MT-kinetochore 
linker stiffness
$c=\SI{20}{\pico\newton\per\micro\meter}$ and 20--25 MTs per kinetochore,
 we get
periods of 206--$\SI{258}{s}$ and 103--$\SI{129}{s}$ for kinetochore and
breathing oscillations, respectively.
Our approach reproduced the
frequency doubling of breathing compared to single kinetochore oscillations,
observed in the experiment \cite{wan2012}.
  Both in the model and in the experiment this
doubling originates from the different velocities of AP- and P-moving
kinetochores, which ensure that a P-to-AP switch
($3/3'$ in figure~\ref{fig:osc})
always follows an AP-to-P switch
of the same kinetochore ($1/1'$ in figure~\ref{fig:osc}).
 In the model the velocity  difference is, however, much larger. As a
consequence, in our model with 20--25 MTs an AP-to-P switch follows
96--$\SI{119}{s}$ after a P-to-AP switch of the sister kinetochore, which is
\SI{93}{\percent} of a breathing period, whereas in PtK2 cells a mean interval
of merely \SI{6}{s} has been measured~\cite{dumont2012}.
  In other words, in our model, most of the time 
both kinetochores move towards each other in AP-direction
(state 0 in figure~\ref{fig:osc}),
whereas in the experiment, mostly one kinetochore moves in P-
while the trailing sister is moving in AP-direction
(state $2/2'$ in figure~\ref{fig:osc}).
In our model,
different AP- and P-velocities are based on the fact that the MT shrinkage is
much faster than growth.  The model parameters for MT dynamics were taken from
experimental measurements with yeast kinetochores~\cite{akiyoshi2010}, which,
however, are distinct from metazoan kinetochores in two main points: firstly,
they can only attach to one MT~\cite{winey1995};  secondly, the Ndc80 fibrils
are connected to MTs via ring-like Dam1 complexes, which do not appear in
metazoan cells~\cite{cheeseman2014}.  
We show in  section\ \ref{sec:ptk1}
  that this discrepancy can be eliminated
  by adjusting some MT parameters
and, moreover, the model can reproduce
kinetochore oscillations in PtK1 cells quantitatively.

In experiments with HeLa cells
Jaqaman {\it et al.}~\cite{jaqaman2010} observed 
an increase of oscillation amplitudes and periods
when they weakened the cohesin bond.
In our model, a smaller cohesin stiffness $c_\mathrm{k}$ has the same two effects
as the inter-kinetochore distance has to be larger to reach the turning points
$F_\mathrm{min}$ and $F_\mathrm{max}$ of the hysteresis loop,
and the phase space velocity 
$\dot{F}_\mathrm{kk} = c_\text{k} \left(v_\mathrm{k,r}+v_\mathrm{k,l}\right)$
and, therefore, the frequencies
are proportional to $c_\mathrm{k}$.

Our analytical approach also allowed us to go beyond the results of
\cite{banigan2015} and 
 quantify constraints on the linker stiffness $c$ and the MT number
for occurrence of bistability in the one-sided model and for the occurrence of
oscillations in the full model.  
We found that bistability requires linker stiffnesses above 
$c_\mathrm{bist}\simeq \SI{8}{\pico\newton\per\micro\meter}$. 
Bistability is, however, not sufficient for oscillations.
Our phase
space interpretation showed that bistability only leads to directional
instability if the two branches of the force-velocity relation 
are also separated by the zero velocity line.
This condition quantifies
 the oscillatory regime  in the parameter plane
of $c$ and $M$. 
We predict that oscillations should only be observable if the MT-kinetochore 
linker stiffness is above 
$c_\mathrm{osc} \simeq \SI{16}{\pico\newton\per\micro\meter}$. 
Our model can thus provide 
 additional information on the MT-kinetochore linkers whose 
molecular nature is unknown up to now. 
Several Ndc80 fibrils, which cooperatively bind to the MT, are
 an important part of the MT-kinetochore link and 
the stiffness of this Ndc80 link has been determined
recently using optical trap measurements \cite{Volkov2018}.
These experiments found stiffnesses above 
$\sim \SI{20}{\pico\newton\per\micro\meter}$, which are compatible 
with our bounds. Moreover, they found a stiffening of the link under force,
which could be included in our model in future work.

The derivation of the lower bound for the stiffness for the 
occurrence of oscillations is based on the occurrence of 
a new zero AP-velocity fixed point in the 
force-velocity diagram of the kinetochores, which suppresses
oscillations  upon decreasing the stiffness. 
Also the  influence of poleward flux to the
system could be analyzed by a fixed point analysis of the 
force-velocity diagram. 
 Since poleward MT flux shifts the force-velocity 
towards smaller AP-velocities of the kinetochore,
 the upper branch may cross zero velocity establishing again a 
 zero velocity fixed point suppressing oscillations. This explains 
why  high
flux velocities suppress directional instability and rationalizes 
the correlation between kinetochore oscillations and poleward flux observed in
several cells (table~\ref{tab:flux}).
  It has been observed in newt lung cells that oscillations
are occasionally ($\SI{11}{\percent}$ of time) interrupted by phases
in which the kinetochores pause their motion~\cite{skibbens1993}
analogously to resting in the fixed point in our model.
This indicates that the spindle of newt lung cells
operates near the boundary between the oscillatory
and the non-oscillatory regime.

  Also experimental results
  in \cite{deluca2006,deluca2011,zaytsev2014,Deluca2018}
on the effects of phosphorylation of Hec1,
which is part of mammalian Ndc80 complex,
onto kinetochore dynamics 
can be rationalized by our  force-velocity diagram of the kinetochores.
Dephosphorylation
leads to hyper-stable MT-kinetochore attachments,
increases the inter-kinetochore distance,
damps or completely suppresses oscillations,
and lets the kinetochores more often be found in a ``paused state''.
The increase of the inter-kinetochore distance can be explained
with the hyper-stable MT-kinetochore attachments:
in the oscillatory regime, the bistable area of the force-velocity relation
increases if more MTs are attached to the kinetochore
(figure~\ref{fig:xvk}(b));
in the non-oscillatory regime,
the mean distance $\langle \Delta X_\mathrm{k}\rangle$
is a linear function of $M$ (\eref{eqn:dxkmean}).
However, the suppression of
oscillations and the frequent appearance of paused states,
which are both effects of leaving the oscillatory regime in our model,
can not be explained with an increasing number of attached MTs.
Instead, we suggest three additional effects of Hec1 phosphorylation:
Firstly, it is imaginable that Hec1 is a catastrophe factor 
that is activated by phosphorylation, i.e.,
if phosphorylation is suppressed, the catastrophe rate at the kinetochore
$\omega_\mathrm{c}^\mathrm{kin}$ decreases.
Secondly, phosphorylation of Hec1 could stiffen the Ndc80 complex
so that dephosphorylation suppresses oscillations
by decreasing the linker stiffness $c$.
Since the stiffness of the Ndc80 complex
has been measured in a recent experiment~\cite{Volkov2018},
this second option might be testable.
The third possible explanation is based on the observation
of Umbreit {\it et al.}~\cite{umbreit2012}
that phosphorylation of Hec1 suppresses rescues.
Following the argumentation in Sec.~\ref{sec:MFtheory},
we conclude that an decreased rescue rate
has a similar effect as an increase of the linker stiffness:
Since the exponent $\alpha_-$ that defines the leading
order of $p(x)$ near $x_\mathrm{min}$ is a linear function
of $\omega_\mathrm{r}^0$ ($\alpha_-+1\propto\omega_\mathrm{r}^0$,
see supplementary material),
the probability density $p(x)$ becomes sharper for negative
kinetochore velocities if rescue is suppressed,
finally leading to a bistable master curve
that allows for oscillations.
In \cite{zaytsev2014}, besides suppression,
Hec1 phosphorylation has also been enforced on up to four sites.
As a result, the number of attached MTs 
and the periods of kinetochore oscillations decreased,
which is consistent with our model (figure~\ref{fig:unconfined}(c)).
Moreover, kinetochore oscillations were supported
but became more erratic
just like in our model, where kinetochore motion is more
fluctuative if less MTs are attached (figure~\ref{fig:unconfined}(ab)).
This experimental result reinforces our point of view
that regular kinetochore oscillations are an emergent phenomenon
that results from the collective behavior of 
stochastic MT dynamics.

Furthermore, we added linearly distributed PEFs,
which depend on the absolute kinetochore positions.
Their main effect is a phase shift between the sister kinetochores
in their phase space trajectories,
which leads to regularly alternating kinetochore oscillations
and, finally, forces the kinetochores to stay near the spindle equator.
Consistently, experimental results show
that a proper formation of the metaphase plate is not assured 
when PEFs are suppressed~\cite{levesque2001}.
Since the PEFs do not affect the master curves and phase space diagrams,
deterministically, they never completely suppress oscillations
but only reduce their amplitude and increase their frequency,
while the kinetochore velocities $v_\mathrm{k}$ are unchanged.
This is consistent with experiments of Ke {\it et al.}~\cite{ke2009},
who observed an increase in amplitude but no influence on
the occurrence of oscillations and the velocity of chromosomes
after severing the chromosome arms
and thereby weakening the PEFs.
In stochastic simulations, the kinetochore oscillations
are more fluctuative in the presence of PEFs,
see figure~\ref{fig:PEFsimu}.
A similar observation was made in the model of Civelekoglu-Scholey 
{\it et al.}~\cite{civelekoglu2013}.
Moreover, in stochastic simulations, 
sister kinetochores tend to oscillate in antiphase 
and frequency doubling of breathing compared
to single kinetochore oscillations is lost
if PEFs are strong compared to the inter-kinetochore tension
($k\gg c_\mathrm{k}$).
Since to our knowledge 
such antiphase oscillations have not been observed in vivo,
we conclude that the inter-kinetochore tension
is the dominating force for directional instability.

Consistently with experimental observations
in both fission yeast~\cite{mary2015,gergely2016}
and human cells~\cite{mayr2007},
kinesin-8 motors investigated
in the model of Klemm {\it et al.}~\cite{klemm2018}
have a similar centering
effect as the PEFs in our model.
Since fission yeast does not contain chromokinesins~\cite{wood2002},
the Klemm model does not include PEFs, whereas
our model does not include  kinesin-8.
It remains an open question 
whether and how the similar effects of PEFs
and kinesin-8 cooperate if both are present.
As kinesin-8 depolymerizes MTs in a length-dependent
manner~\cite{varga2006,varga2009},
it could be included in our model 
by a catastrophe rate $\omega_\mathrm{c}$
that depends on the  MT length $x_\mathrm{m}$,
While such MT length-dependent catastrophe rates
can easily be implemented
in the stochastic simulations,
they are difficult to include into 
our  mean-field theory,
which is based on solving the FPEs~\eref{eqn:FPG+}
and~\eref{eqn:FPG-} in relative coordinates
rather than absolute MT lengths.

Finally, we lifted the assumption 
that  MTs are able to apply pushing forces on the kinetochores
because experiments suggest that MTs only exert tensile forces
\cite{waters1996,khodjakov1996}.
Therefore, we confined MT growth at the kinetochore
by  catastrophe-triggering boundary conditions.
The catastrophe rate for a MT at the kinetochore
$\omega_\mathrm{c}^\mathrm{kin}$
can, in principle, 
 range from the  force-free MT catastrophe rate $\omega^0_\mathrm{c}$,
which is realistic in the absence of any catastrophe promoting 
proteins up to infinity if a catastrophe is immediately triggered. 
In the presence of the  centromere-associated regulating protein MCAK
increased catastrophe rates
$\omega_\mathrm{c}^\mathrm{kin}=7\omega^0_\mathrm{c}$
are expected \cite{newton2004}.
We found that {\it both} the linker stiffness $c$ {\it and} 
the catastrophe rate $\omega_\mathrm{c}^\mathrm{kin}$ at the kinetochore 
have to be sufficiently large to obtain bistability and  oscillations.
We find, in particular,  that the  force-free MT catastrophe rate
is not sufficient to lead to oscillations, which shows 
that catastrophe-promoting proteins are essential to induce 
oscillations. 
In the presence of PEFs, oscillations can be recovered also for 
relatively  small catastrophe rates:
For $\omega_\mathrm{c}^\mathrm{kin}/\omega^0_\mathrm{c} \sim 5$, we found 
no oscillations in the absence of PEFs; for 
 $\omega_\mathrm{c}^\mathrm{kin}/\omega^0_\mathrm{c} <2$ we found no 
oscillations at all.
Moreover, the linker stiffness has to be sufficiently high 
to give  linker extensions below 
 $\SI{100}{\nano\meter}$ compatible with experimental results
\cite{Wei2005,Wang2008,Maresca2009,dumont2012}.
For $\omega_\mathrm{c}^\mathrm{kin}/\omega^0_\mathrm{c}=20$
and  a linker stiffness of $c=\SI{20}{\pico\newton\per\micro\meter}$,
we found realistic behavior. 
Our results can explain experimental observations in 
\cite{kline2004},
where PtK2-cells were observed under depletion
 of centromeric MCAK, which decreases $\omega_\mathrm{c}^\mathrm{kin}$.
Then, in accordance to our results (see  figure\ \ref{fig:conf_wckin}(cd)), 
the  turning point  $F_\mathrm{max}$ of the hysteresis loop
decreases. As a result 
the oscillation frequency increases and the mean centromere stretch decreases,
while the ``motility rates'', i.e., the velocities  do not change.

Kinetochore motion in the non-oscillatory regime can be described as
fluctuations around a fixed point with constant 
inter-kinetochore distance.  This is exactly the behavior of peripheral
kinetochores in PtK1 cells~\cite{wan2012,civelekoglu2013},
while the central kinetochores
do exhibit directional instability.  Civelekoglu-Scholey 
{\it et al.}~\cite{civelekoglu2013} explained this dichotomy with different
distributions of polar ejection forces in the center and the periphery of the
metaphase plate. 
However, the model kinetochore trajectories in the presence of strong PEFs,
which they declare to be 
representative for the motion of peripheral kinetochores
(figure~6C in~\cite{civelekoglu2013}),
still have a regular oscillating shape
with only a reduced amplitude and an increased frequency,
in agreement with the results of our model in figure~\ref{fig:PEFsimu}.
The experimental trajectories for peripheral kinetochores
from\ \cite{wan2012,civelekoglu2013}, on the other hand,
are very fluctuative, hardly show any regular oscillations,
and are very similar to the trajectories
in the non-oscillatory regime of our model.
For a clear characterization of the experimentally measured
motion of peripheral kinetochores
as either stochastic fluctuations or regular oscillations
its representation in Fourier space would be helpful
as already provided for the central kinetochores
by Wan {\it et al.}~\cite{wan2012} and as provided
  in  figure~\ref{fig:PEFsimu} for our model.
If the Fourier transforms do not have any distinct peaks,
differences in PEFs are ruled out as a possible explanation
for the dichotomy in PtK1 cells
according to both our model
and the one of Civelekoglu-Scholey {\it et al.}

Instead, our results suggest differences in linker stiffness
or catastrophe promotion as  reasons for the dichotomy. 
 For instance, 
less Ndc80 complexes could  participate in peripheral MT-kinetochore links 
resulting in reduced linker stiffness and 
non-oscillatory behavior. 
Also  a non-uniform MCAK distribution that 
 decreases radially towards the periphery of the metaphase plate
could reduce   $\omega_\mathrm{c}^\mathrm{kin}$  and suppress 
oscillations 
 of   peripheral kinetochores.
Differences in poleward flux might be another
possible explanation for the dichotomy according to our results.
However, Cameron {\it et al.}~\cite{cameron2006} observed
that the flux velocities in PtK1 cells do not depend
on the chromosome to which a MT is attached.

In conclusion, the minimal model can rationalize a number of experimental
observations. Particularly interesting are 
constraints on the MT-kinetochore linker stiffness that are 
compatible with recent optical trap measurements \cite{Volkov2018}. 
The predicted responses to the most 
relevant parameter changes  are summarized in table\ \ref{tab:summary}
and suggest 
further systematic perturbation experiments, for example,
by promoting catastrophes at the kinetochore.

\begin{table}[!ht]
	\caption{Summary.
		Effect of an increase of the parameter in the first column 
		on occurrence, frequency, and amplitude of kinetochore oscillations.
	}
	\label{tab:summary}
	\footnotesize
	\begin{tabular}{lllllm{0.23\linewidth}}
		\br
		parameter & symbol & occurrence & frequency & amplitude & additional effects \\
		\mr
		linker stiffness & $c$ & stimulation & decrease & increase & decrease of inter-kinetochore distance in non-oscillatory regime\\
		poleward flux & $v_\mathrm{f}$ & suppression & decrease & none &\\
		polar ejection forces & $k$ & none & increase & decrease & PEFs force kinetochores to oscillate alternately and to stay near the spindle equator \\
		catastrophe rate of stalled MTs & $\omega_\mathrm{c}^\mathrm{kin}$ & stimulation & decrease & increase &\\
		cohesin bond stiffness & $c_\mathrm{k}$ & none & increase &  decrease & \\
		MT number & $M$ & (stimulation) & decrease &  increase & \\
		\br
	\end{tabular}
\end{table} 

\ack
{We acknowledge support by the Deutsche Forschungsgemeinschaft (grant
No. \mbox{KI 662/9-1}).
We acknowledge financial support by Deutsche Forschungsgemeinschaft and TU Dortmund
University within the funding programme Open Access Publishing.
}

\section*{References}
\bibliography{MT_kinetochore_dynamics}

\end{document}